\documentclass[pdflatex,sn-basic]{sn-jnl}
\usepackage{titlesec}
\usepackage{natbib}
\usepackage{enumitem}
\usepackage{tabularx}
\usepackage{booktabs}
\usepackage{bookmark}
\usepackage{graphicx}%
\usepackage{multirow}%
\usepackage{amsmath,amssymb,amsfonts}%
\usepackage{amsthm}%
\usepackage{hyperref}
\usepackage{bm}
\usepackage{dsfont}
\usepackage{mathrsfs}%
\usepackage[title]{appendix}%
\usepackage{xcolor}%
\usepackage{textcomp}%
\usepackage{manyfoot}%
\usepackage{booktabs}%
\usepackage{algorithm}%
\usepackage{algorithmic}
\usepackage{listings}%
\usepackage{bookmark}
\usepackage{anyfontsize}
\usepackage{tikz}
\usepackage{tikz-3dplot}
\usepackage{colortbl} 
\definecolor{lightgray}{gray}{0.9}
\usepackage{makecell} 
\usepackage{rotating}
\usepackage{multirow}
\usepackage{adjustbox} 
\usepackage{rotating}
\usepackage{pdflscape}

\renewenvironment{table}[1][]%
{\tableorg[#1]%
\tablebodyfont%
\renewcommand\footnotetext[2][]{{\removelastskip\vskip3pt%
\let\tablebodyfont\tablefootnotefont%
\hskip0pt\if!##1!\else{\smash{$^{##1}$}}\fi##2\par}}%
}{\endtableorg}

\usepackage{float}

\DeclareMathOperator*{\argmin}{arg\,min}

\allowdisplaybreaks

\newtheorem{assumption}{Assumption}

\begin{document}

\title[Article Title]{Gradient Boosting for Spatial Panel Models with Random and Fixed Effects}

\author*[1]{\fnm{Michael} \sur{Balzer}}\email{michael.balzer@uni-bielefeld.de}

\author[1]{\fnm{Adhen} \sur{Benlahlou}}\email{adhen.benlahlou@uni-bielefeld.de}

\affil[1]{\orgdiv{Bielefeld University}, \orgname{Center for Mathematical Economics}, \orgaddress{\street{Universitätsstraße 25}, \city{Bielefeld}, \postcode{33615}, \state{NW}, \country{Germany}}}

\abstract{Due to the increase in data availability in urban and regional studies, various spatial panel models have emerged to model spatial panel data, which exhibit spatial patterns and spatial dependencies between observations across time. Although estimation is usually based on maximum likelihood or generalized method of moments, these methods may fail to yield unique solutions if researchers are faced with high-dimensional settings. This article proposes a model-based gradient boosting algorithm, which enables estimation with interpretable results that is feasible in low- and high-dimensional settings. Due to its modular nature, the flexible model-based gradient boosting algorithm is suitable for a variety of spatial panel models, which can include random and fixed effects. The general framework also enables data-driven model and variable selection as well as implicit regularization where the bias-variance trade-off is controlled for, thereby enhancing accuracy of prediction on out-of-sample spatial panel data. Monte Carlo experiments concerned with the performance of estimation and variable selection confirm proper functionality in low- and high-dimensional settings while real-world applications including non-life insurance in Italian districts, rice production in Indonesian farms and life expectancy in German districts illustrate the potential application.}


\maketitle

\section{Introduction} \label{sec:intro}
Economic outcomes frequently exhibit spatial and temporal dependencies, so that changes in one geographical location can influence nearby locations and evolve over time. Linear panel data models usually ignore this so-called spatial autocorrelation, which can lead to biased estimates and invalid inference. To account for spatial autocorrelation in data, linear panel data models have been extended to spatial panel data models, explicitly incorporating the spatio-temporal dependence structure. One common approach assumes that spatial autocorrelation enters the model only through the error term, allowing for spatially autocorrelated location-specific random effects and spatially autocorrelated remainder disturbances. Such a model is 
referred to as the generalized spatial panel model with error components (GSPECM) \citep{baltagi2008, baltagi2013}.

Estimation of the GSPECM is typically conducted using maximum likelihood (ML) or generalized method of moments (GMM). These methods are well behaved in low-dimensional settings where the number of variables is small compared to the number of spatial units. However, in many modern applications, the number of potential variables is large while the number of spatial units grows slowly or remains fixed. In such high-dimensional settings, conventional ML and GMM estimators can be ill-defined or fail to yield unique solutions, making model and variable selection essential. Since standard ML and GMM estimators do not provide an inherent variable selection mechanisms, recent research has extended penalized estimation techniques to spatial panel models, including penalized quasi-maximum likelihood and the least absolute shrinkage and selection operator \citep{xia2023, liu2024}.

Alternatively, model-based gradient boosting has been extended to classical linear panel models \citep{hao2024}. This iterative algorithm estimates the model by sequentially fitting base-learners to the negative gradient of a prespecified loss function, where each base-learner represents the effect of a variable and the loss function corresponds to the statistical model of interest. At each iteration, only the best-performing base-learner is selected and a small portion of its contribution is added to the current linear predictor. Early stopping provides a data-driven approach to model and variable selection, producing interpretable results at every iteration. Because of its modular and iterative nature, model-based gradient boosting remains computationally feasible even in high-dimensional settings \citep{bühlmann2006, bühlmann2007, mayr2014}.

Although model-based gradient boosting has been applied to cross-sectional spatial regression models with autoregressive disturbances, no prior work appears to have addressed a potential extension of model-based gradient boosting for the GSPECM. This article therefore develops model-based gradient boosting for the GSPECM, building on the algorithm proposed by \cite{balzer2025}. The algorithm is implemented in the \texttt{mboost} package, which supports gradient boosting for generalized linear, additive, and interaction models of potentially high dimensionality in R \citep{bühlmann2007, hothorn2010, hofner2014, hofner2015, R}. Its performance is assessed via extensive Monte Carlo experiments in both low- and high-dimensional settings, focusing on estimation accuracy and variable selection. The approach is further illustrated through three empirical applications, which are concerned with the determinants of non-life insurance in Italian provinces, rice production on Indonesian farms and life expectancy in German districts.

The rest of the article is organized as follows. Section \ref{sec:meth} introduces the GSPECM, the associated estimator as well as the model-based gradient boosting algorithm. Monte Carlo experiments are reported in Section \ref{sec:sim}. Finally, empirical illustrations are presented in Section \ref{sec:illu} while Section \ref{sec:con} concludes the article.

\section{Methodology} \label{sec:meth}
\subsection{Spatial Panel Models with Error Components}
Let $N \in \mathbb{N}$ be the number of observations and $T \in \mathbb{N}$ the number of time periods in a spatial data set. For each location $i \in \{1, \dots, N\}$ and time period $t \in \{1,\dots,T\}$, consider following GSPECM
\begin{equation} 
\begin{aligned}
\bm{y}_t &= \bm{X}_t \bm{\beta} + \bm{W}\bm{X}_t \bm{\theta} + \bm{v}_t \\
\bm{v}_t &= \bm{u}_1 + \bm{u}_{2t} \\
\bm{u}_1 &= \rho_1  \bm{W} \bm{u}_1 + \bm{\mu} \\
\bm{u}_{2t} &= \rho_2  \bm{W} \bm{u}_{2t} + \bm{\epsilon}_t
\end{aligned}
\end{equation}
where $\bm{y}_t$ is the $N \times 1$ vector of observations, $\bm{X}_t$ is the $N \times P$  non-stochastic design matrix of $P \in \mathbb{N}$ exogenous variables, $\bm{\beta}$ is the time-invariant $P \times 1$ coefficient vector and $\bm{v}_t$ is the $N \times 1$ vector of disturbances at time $t$. The spatial autocorrelation enters the model in two ways in which the $N \times N$ spatial weight matrix $\bm{W}$ that captures spatial connections between observations plays a crucial role. A $N \times P$ non-stochastic design matrix of spatial lags of exogenous variables $\bm{W}\bm{X}_t$ and corresponding time-invariant $P \times 1$ coefficient vector $\bm{\theta}$ capture spatial autocorrelation in the explanatory variables while the disturbance term $\bm{v}_t$ follows an error components model with spatial autocorrelation parameters.\footnote{Note that a spatial dependence structure in data can also be accounted for by additionally incorporating higher-order spatial lags of exogenous variables $\bm{W}\bm{W}\bm{X}_t$ or even $\bm{W}\bm{W}\bm{W}\bm{X}_t$ as regressors.} Particularly, $\bm{u}_1$ is the $N \times 1$ time-invariant location-specific vector of disturbances, which includes the $N \times 1$ vector of location-specific random effects $\bm{\mu}$ and $\bm{u}_{2t}$ is the $N \times 1$ time-varying vector of remainder disturbances, which includes the $N \times 1$ vector of idiosyncratic random innovations $\bm{\epsilon}_t$. Elements of $\bm{\mu}$ and $\bm{\epsilon}$ are independently and identically distributed according to $\bm{\mu} \sim \mathcal{N}(0,\sigma_{\bm{\mu}}^2)$ across $i$ and $\bm{\epsilon} \sim \mathcal{N}(0,\sigma_{\bm{\epsilon}}^2)$ across $i$ and $t$, respectively. Additionally, both disturbances terms are allowed to depend on the spatial weight matrix $\bm{W}$ and spatial autocorrelation parameters $\rho_1$ and $\rho_2$, which capture spatial autocorrelation in the disturbance term.

Stacking the cross-sections together produces the GSPECM in matrix form
\begin{equation}
\begin{aligned}
\bm{y} &= \bm{X} \bm{\beta} + (\bm{I}_T \otimes \bm{W})\bm{X} \bm{\theta} + \bm{v} \\
\bm{v} &= \bm{L}_{\bm{\mu}} \bm{u}_1 + \bm{u}_{2} \\
\bm{u}_1 &= \rho_1  \bm{W} \bm{u}_1 + \bm{\mu} \\
\bm{u}_{2} &= \rho_2  (\bm{I}_T \otimes \bm{W}) \bm{u}_{2} + \bm{\epsilon}
\end{aligned}
\end{equation}
where $\bm{y} = \left(\bm{y}_1^\prime, \dots, \bm{y}_T^\prime\right)^\prime$, $\bm{X} = \left[\bm{X}_1^\prime, \dots, \bm{X}_T^\prime\right]^\prime$, $\bm{v} = \left(\bm{v}_1^\prime, \dots, \bm{v}_T^\prime\right)^\prime$, $\bm{u}_2 = \left(\bm{u}_{21}^\prime, \dots, \bm{u}_{2T}^\prime\right)^\prime$ and $\bm{\epsilon} = \left(\bm{\epsilon}_{1}^\prime, \dots, \bm{\epsilon}_{T}^\prime\right)^\prime$. Consequently, $\bm{I}_T$ is the $T \times T$ identity matrix while the time-invariant location-specific disturbances $\bm{u}_1$ are repeated in all times periods via the $NT \times N$ selector matrix $\bm{L}_{\bm{\mu}} = \bm{\iota}_T \otimes \bm{I}_N$, in which $\bm{\iota}_T$ is the $T \times 1$ vector of ones and $\bm{I}_N$ is the $N \times N$ identity matrix. Proper estimation of the spatial autocorrelation parameters, the coefficients of the exogenous variables, and their corresponding spatial lags is ensured by imposing the following regularity conditions.

\begin{assumption} \label{ass:hom}
\par\noindent
\begin{enumerate}[label=(\alph*)]
    \item For all $N \ge 1$, $1 \le i \le N$, and $1 \le t \le T$, the idiosyncratic random innovations $\epsilon_{it}$ are identically and independently distributed with mean zero and variance $\sigma_{\bm{\epsilon}}^2$ where $0 < \sigma_{\bm{\epsilon}}^2 < b_{\bm{\epsilon}} < \infty$ and possess finite fourth moments. 
    
    \item For all $N \ge 1$ and $1 \le i \le N$, the location-specific effects $\mu_{i}$ are identically and independently distributed with mean zero and variance $\sigma_{\bm{\mu}}^2$ where $0 < \sigma_{\bm{\mu}}^2 < b_{\bm{\mu}} < \infty$ and possess finite fourth moments.
    
    \item The collections of random variables $\bm{\mu}$ and $\bm{\epsilon}$ are mutually independent.
\end{enumerate}
\end{assumption}

\begin{assumption}
\par\noindent
\begin{enumerate}[label=(\alph*)]
    \item The spatial weight matrix $\bm{W}$ has no self-loops such that the diagonal entries satisfy $w_{ii} = 0$. The off-diagonal entries satisfy $w_{ij} = O\left(\frac{1}{H}\right)$ where $\frac{H}{N} \to 0$.
    
    \item For any $|\rho_1| < 1$ and $|\rho_2| < 1$, the matrices $\bm{I}_N - \rho_1 \bm{W}$ and $\bm{I}_N - \rho_2 \bm{W}$ exist and are non-singular.
    
    \item The row and column sums of $\bm{W}$ and $\left(\bm{I}_N - \rho_1 \bm{W}\right)^{-1}$ and $\left(\bm{I}_N - \rho_2 \bm{W}\right)^{-1}$  are uniformly bounded in absolute value.
\end{enumerate}
\end{assumption}

The GSPECM thus generalizes a variety of spatial panel models discussed in the literature. If $\rho_1 = \rho_2$, the GSPECM reduces to the spatial random effects model of \cite{kapoor2007} (KKP), whereas setting $\rho_1 = 0$ yields the spatial random effects model of \cite{anselin1988} (ANS). Both models are widely employed in applied work. If the spatial autocorrelation parameters are eliminated entirely $\rho_1 = \rho_2 = 0$, the GSPECM simplifies to the random effects panel data model \citep{baltagi2008, mutl2011, baltagi2013, baltagi2016}.

\subsubsection{Random and Fixed Effects Specification}
Combining the exogenous variables with their spatial lags into the design matrix $\bm{Z} = \left[\bm{X}, (\bm{I}_T \otimes \bm{W})\bm{X}\right]$ and the corresponding coefficients into $\bm{\delta} = (\bm{\bm{\beta}^{\prime}, \bm{\theta}}^{\prime})^{\prime}$ yields the linear predictor $\bm{\eta} = \bm{Z} \bm{\delta}$ and thereby the GSPECM in compact form as 
\begin{equation} \label{eq:gspecm}
     \bm{y} = \bm{\eta} + \bm{v}, \quad 
    \mathbb{E}(\bm{v}) = \bm{0}, \ \text{Var}(\bm{v}) = \bm{\Omega}.
\end{equation}
The following regularity condition is imposed on $\bm{Z}$.
\begin{assumption} 
    The matrix $\bm{Z}$ has full column rank. Specifically, the limit $\lim_{NT \to \infty} \bm{Z}^{\prime} \bm{Z}$ exists, is non-singular and the elements of $\bm{Z}$ are uniformly bounded in absolute value.
\end{assumption}

The loss function associated with Equation \eqref{eq:gspecm} is the squared Mahalanobis distance of the disturbance term, arising from the generalized least squares objective function
\begin{equation}
    f(\bm{y}, \bm{\eta}, \bm{\Omega}) = 
    (\bm{y} - \bm{\eta})^{\prime} \bm{\Omega}^{-1} (\bm{y} - \bm{\eta})
\end{equation}
such that the negative gradient of the loss function with respect to the linear predictor $\bm{\eta}$ is
\begin{equation} 
    -\frac{\partial}{\partial \bm{\eta}} f(\bm{y}, \bm{\eta}, \bm{\Omega}) = 
    2 \bm{\Omega}^{-1} (\bm{y} - \bm{\eta}).
\end{equation}
Let $\bm{A} = \left(\bm{I}_N - \rho_1 \bm{W}\right)$ and $\bm{B} = \left(\bm{I}_N - \rho_2 \bm{W}\right)$, then the distributions of the disturbance terms are given by
\begin{align}
    \bm{u}_1 &= \bm{A}^{-1} \bm{\mu} \sim \mathcal{N}(0, \sigma_{\bm{\mu}}^2 (\bm{A}^\prime \bm{A})^{-1})\\
    \bm{u}_2 &= \left(\bm{I}_T \otimes \bm{B}^{-1} \right) \bm{\epsilon} \sim \mathcal{N}(0, \sigma_{\bm{\epsilon}}^2 (\bm{I}_T \otimes (\bm{B}^\prime \bm{B})^{-1})).
\end{align}
Let $\bm{E}_T = \bm{I}_T - \bm{\bar J}_T$ and $\bm{\bar J}_T = \frac{\bm{J}_T}{T}$, where $\bm{\bar J}_T$ is the $T \times T$ averaging matrix and $\bm{J}_T$ is the $T \times T$ matrix of ones. Thus, the concentration matrix of the GSPECM is given by
\begin{equation}
    \bm{\Omega}^{-1} = \bm{\bar J}_T \otimes \left[T\sigma_{\bm{\mu}}^2 (\bm{A}^\prime \bm{A})^{-1} + \sigma_{\bm{\epsilon}}^2(\bm{B}^\prime \bm{B})^{-1}\right]^{-1} + \frac{1}{\sigma_{\bm{\epsilon}}^2} \left[\bm{E}_T \otimes (\bm{B}^\prime \bm{B})\right],
\end{equation}
yielding all necessary ingredients for the model-based gradient boosting algorithm in the random effects specification.

The random effects specification relies on the critical assumption that the unobserved time-invariant location-specific disturbances $\bm{u}_1$ are uncorrelated with the entries of $\bm{Z}$. If the assumption is violated, then $\mathbb{E}(u_{1i} | Z_{it}) \neq 0$ such that model-based gradient boosting with the random effects specification is inappropriate. Then, the fixed effects specification can be utilized instead, which has an identical representation for the GSPECM as well as the ANS and KKP random effects model \citep{lee2010}. Specifically, let $\bm{Q} = \bm{E}_T \otimes \bm{I}_N$ denote the within transformation. This can be combined with a Cochrane--Orcutt--type spatial transformation using $(\bm{I}_T \otimes \bm{B})$, such that $\bm{Q}(\bm{I}_T \otimes \bm{B}) = (\bm{E}_T \otimes  \bm{B})$ 
to obtain
\begin{equation} \label{eq:gspecmFE}
    (\bm{E}_T \otimes  \bm{B})\bm{y} = (\bm{E}_T \otimes  \bm{B})\bm{\eta} + \bm{Q}\bm{\epsilon} \quad \mathbb{E}(\bm{\epsilon}) = 0, \ \text{Var}(\bm{\epsilon}) = \sigma_{\bm{\epsilon}}^2 \bm{I}_{NT}.
\end{equation}
The loss function associated with Equation \eqref{eq:gspecmFE} is the squared Mahalanobis distance of the disturbance term, arising from the generalized least squares objective function
\begin{equation}
    f(\bm{y}, \bm{\eta}, \bm{\Psi}) = 
    (\bm{y} - \bm{\eta})^{\prime} \bm{\Psi}^{-1} (\bm{y} - \bm{\eta})
\end{equation}
such that the negative gradient of the loss function with respect to the linear predictor $\bm{\eta}$ is
\begin{equation} 
    -\frac{\partial}{\partial \bm{\eta}} f(\bm{y}, \bm{\eta}, \bm{\Psi}) = 
    2 \bm{\Psi}^{-1} (\bm{y} - \bm{\eta}).
\end{equation}
In this case, the variance-covariance matrix of the idiosyncratic random innovations $\bm{\epsilon}$ is adjusted after inverting by the within and Cochrane–Orcutt type spatial transformation and the concentration matrix is thus given by
\begin{equation} 
    \bm{\Psi}^{-1} = \frac{1}{\sigma_{\bm{\epsilon}}^2} (\bm{E}_T \otimes (\bm{B}^\prime \bm{B})),  
\end{equation}
yielding all necessary ingredients for the model-based gradient boosting algorithm in the fixed effects specification \citep{baltagi2016, balzer2025}.\footnote{Note that the variance-covariance matrix before inverting is also adjusted by the within and Cochrane–Orcutt type spatial transformation and given by $\bm{\Psi} = \sigma_{\bm{\epsilon}}^2 (\bm{E}_T \otimes (\bm{B}^\prime \bm{B}))$.}

\subsubsection{Generalized Method of Moments Estimation}
The feasibility of model-based gradient boosting relies on the assumption that the spatial autocorrelation parameters $\rho_1$, $\rho_2$ and the variances of the location-specific random effects $\sigma_{\bm{\mu}}^2$ as well as the idiosyncratic random innovations $\sigma_{\bm{\epsilon}}^2$ are simultaneously apriori known. However, in real-world application settings, $\rho_1$, $\rho_2$, $\sigma_{\bm{\mu}}^2$ and $\sigma_{\bm{\epsilon}}^2$ are unknown, making the evaluation of $\bm{\Omega}$ and $\bm{\Psi}$, which appear in the squared Mahalanobis distance and the negative gradient, infeasible.

To obtain a feasible model-based gradient boosting for the GSPECM, $\rho_1$, $\rho_2$, $\sigma_{\bm{\mu}}^2$ and $\sigma_{\bm{\epsilon}}^2$ have to be replaced by corresponding estimates $\hat{\rho}_1$, $\hat{\rho}_2$, $\hat{\sigma}_{\bm{\mu}}^2$ and $\hat{\sigma}_{\bm{\epsilon}}^2$, which can be conveniently computed via GMM estimators. More specifically, define $\bm{\tilde{v}} = \bm{y} -\bm{Z}\bm{\tilde{\delta}}$ as the predictors of $\bm{v}$ based on a consistent estimator $\bm{\tilde{\delta}}$, ignoring random effects and spatial autocorrelation. Furthermore, define $\bm{\bar{u}}_2 = \left(\bm{I}_T \otimes \bm{W}\right) \bm{u}_2$ and $\bm{\bar{\bar{u}}}_2 = \left(\bm{I}_T \otimes \bm{W}\right) \bm{\bar{u}}_2$ such that $\bm{\epsilon} = \bm{u}_2 - \rho_2 \bm{\bar{u}}_2$ and $\bm{\bar{\epsilon}} = \bm{\bar{u}}_2 - \rho_2 \bm{\bar{\bar{u}}}_2$, then GMM estimators of $\rho_2$ and $\sigma_{\bm{\epsilon}}^2$ are based on following three moment conditions
\begin{align}\label{eq:mom1}
\frac{1}{N(T-1)} \mathbb{E}\!\left(\bm{\epsilon}^\prime \bm{Q} \bm{\epsilon} \right)
&= \sigma_{\bm{\epsilon}}^2, \\
\frac{1}{N(T-1)}\mathbb{E}\!\left(\bm{\bar{\epsilon}}^\prime \bm{Q} \bm{\bar{\epsilon}} \right)
&= \sigma_{\bm{\epsilon}}^2 \frac{1}{N}\operatorname{tr}\!\left(\bm{W}^\prime \bm{W}\right), \\
\frac{1}{N(T-1)}\mathbb{E}\!\left(\bm{\bar{\epsilon}}^\prime \bm{Q} \bm{\epsilon} \right)
&= 0.
\end{align}
By construction, $\bm{Q} (\bm{\iota}_T \otimes \bm{u}_1) = 0$ and thus it follows that $\bm{v}^\prime \bm{Q} \bm{v} = \bm{u}_2^\prime \bm{Q} \bm{u}_2$, $\bm{\bar{v}}^\prime \bm{Q} \bm{\bar{v}} = \bm{\bar{u}}_2^\prime \bm{Q} \bm{\bar{u}}_2$ and $\bm{\bar{v}}^\prime \bm{Q} \bm{v} = \bm{\bar{u}}_2^\prime \bm{Q} \bm{u}_2$ such that a system of three equations can be obtained as
\begin{equation} \label{eq:sol1}
    \bm{\Gamma_0}[\rho_2, \rho_2^2, \sigma_{\bm{\epsilon}}^2]^{\prime} - \bm{\gamma_0} = 0.
\end{equation}
The expressions for $\bm{\Gamma}_0$ and $\bm{\gamma}_0$ are given by 
\begin{equation*}
\begin{aligned}
\bm{\Gamma_0} &=
\begin{bmatrix}
\frac{2}{N(T-1)} \mathbb{E} (\bm{\bar{v}}^\prime \bm{Q} \bm{v}) 
& -\frac{1}{N(T-1)} \mathbb{E}(\bm{\bar{v}}^\prime \bm{Q} \bm{\bar{v}}) 
& 1 \\
\frac{2}{N(T-1)} \mathbb{E}(\bm{\bar{\bar{v}}}^\prime \bm{Q} \bm{\bar{v}}) 
& -\frac{1}{N(T-1)} \mathbb{E}(\bm{\bar{\bar{v}}}^\prime \bm{Q} \bm{\bar{\bar{v}}}) 
& \frac{1}{N} \operatorname{tr}(\bm{W}^\prime \bm{W}) \\
\frac{1}{N(T-1)} \mathbb{E}\!\left(\bm{\bar{\bar{v}}}^\prime \bm{Q}\bm{v}
+ \bm{\bar{v}}^\prime \bm{Q} \bm{\bar{v}} \right)
& -\frac{1}{N(T-1)} \mathbb{E}(\bm{\bar{\bar{v}}}^\prime \bm{Q}\bm{\bar{v}})
& 0
\end{bmatrix}
\\
\bm{\gamma_0} &=
\begin{bmatrix}
\frac{1}{N(T-1)} \mathbb{E}\!\left(\bm{v}^\prime \bm{Q} \bm{v} \right) \\
\frac{1}{N(T-1)}\mathbb{E}\!\left(\bm{\bar{v}}^\prime \bm{Q} \bm{\bar{v}} \right) \\
\frac{1}{N(T-1)}\mathbb{E}\!\left(\bm{\bar{v}}^\prime \bm{Q} \bm{v} \right)
\end{bmatrix}.
\end{aligned}
\end{equation*}
Replacing theoretical moment matrix and vector in Equation \eqref{eq:sol1} by the corresponding sample moment matrix and vector yields
\begin{equation} \label{eq:est1}
    \bm{G}_0[\rho_2, \rho_2^2, \sigma_{\bm{\epsilon}}^2]^{\prime} - \bm{g}_0 = \bm{\xi}_0(\rho_2, \sigma_{\bm{\epsilon}}^2).
\end{equation}
Define $\bm{\tilde{\bar{v}}} = \left(\bm{I}_T \otimes \bm{W}\right) \bm{\tilde{v}}$ and $\bm{\tilde{\bar{\bar{v}}}} = \left(\bm{I}_T \otimes \bm{W}\right) \bm{\tilde{\bar{v}}}$, then expressions for sample moment matrix $\bm{G}_0$ and vector $\bm{g}_0$ are given by
\begin{equation*}
\begin{aligned}
\bm{G_0} &=
\begin{bmatrix}
\frac{2}{N(T-1)} \bm{\tilde{\bar{v}}}^\prime \bm{Q} \bm{\tilde{v}} 
& -\frac{1}{N(T-1)} \bm{\tilde{\bar{v}}}^\prime \bm{Q} \bm{\tilde{\bar{v}}}
& 1 \\
\frac{2}{N(T-1)} \bm{\tilde{\bar{\bar{v}}}}^\prime \bm{Q} \bm{\tilde{\bar{v}}}
& -\frac{1}{N(T-1)} \bm{\tilde{\bar{\bar{v}}}}^\prime \bm{Q} \bm{\tilde{\bar{\bar{v}}}} 
& \frac{1}{N} \operatorname{tr}(\bm{W}^\prime \bm{W}) \\
\frac{1}{N(T-1)} \left(\bm{\tilde{\bar{\bar{v}}}}^\prime \bm{Q} \bm{\tilde{v}}
+ \bm{\tilde{\bar{v}}} ^\prime \bm{Q} \bm{\tilde{\bar{v}}}  \right)
& -\frac{1}{N(T-1)}\bm{\tilde{\bar{\bar{v}}}}^\prime \bm{Q} \bm{\tilde{\bar{v}}}
& 0
\end{bmatrix}
\\
\bm{g_0} &=
\begin{bmatrix}
\frac{1}{N(T-1)} \bm{\tilde{v}}^\prime \bm{Q} \bm{\tilde{v}} \\
\frac{1}{N(T-1)} \bm{\tilde{\bar{v}}}^\prime \bm{Q} \bm{\tilde{\bar{v}}} \\
\frac{1}{N(T-1)} \bm{\tilde{\bar{v}}}^\prime \bm{Q} \bm{\tilde{v}}
\end{bmatrix}.
\end{aligned}
\end{equation*}

Similarly, define $\bm{\bar{\mu}} = \bm{W} \bm{\mu}$,$\bm{\bar{u}_1} = \bm{W} \bm{u}_1$ and $\bm{\bar{\bar{u}}_1} = \bm{W} \bm{\bar{u}_1}$, then GMM estimators of $\rho_1$ and $\sigma_{\bm{\mu}}^2$ are based on following three moment conditions
\begin{align}\label{eq:mom2}
\frac{1}{N} \mathbb{E}\!\left(\bm{\mu}^\prime \bm{\mu} \right)
&= \sigma_{\bm{\mu}}^2, \\
\frac{1}{N}\mathbb{E}\!\left(\bm{\bar{\mu}}^\prime  \bm{\bar{\mu}} \right)
&= \sigma_{\bm{\mu}}^2 \frac{1}{N}\operatorname{tr}\!\left(\bm{W}^\prime \bm{W}\right), \\
\frac{1}{N}\mathbb{E}\!\left(\bm{\bar{\mu}}^\prime \bm{\mu} \right)
&= 0
\end{align}
such that a system of three equations can be obtained as
\begin{equation} \label{eq:sol2}
    \bm{\Gamma_1}[\rho_1, \rho_1^2, \sigma_{\bm{\mu}}^2]^{\prime} - \bm{\gamma_1} = 0.
\end{equation}
Conveniently, theoretical moment matrix $\bm{\Gamma_1}$ and vector $\bm{\gamma_1}$ can be expressed in terms of $\bm{v}$, $\bm{\bar{v}}$ and $\bm{\bar{\bar{v}}}$ by utilizing $\bm{S} = \left(\bm{\bar{J}_T} - \frac{1}{T-1} \bm{E}_T \right) \otimes \bm{I}_{N}$ as
\begin{equation*}
\begin{aligned}
\bm{\Gamma_1} &=
\begin{bmatrix}
\frac{2}{NT} \mathbb{E} (\bm{\bar{v}}^\prime \bm{S} \bm{v}) 
& -\frac{1}{NT} \mathbb{E}(\bm{\bar{v}}^\prime \bm{S} \bm{\bar{v}}) 
& 1 \\
\frac{2}{NT} \mathbb{E}(\bm{\bar{\bar{v}}}^\prime \bm{S} \bm{\bar{v}}) 
& -\frac{1}{NT} \mathbb{E}(\bm{\bar{\bar{v}}}^\prime \bm{S} \bm{\bar{\bar{v}}}) 
& \frac{1}{N} \operatorname{tr}(\bm{W}^\prime \bm{W}) \\
\frac{1}{NT} \mathbb{E}\!\left(\bm{\bar{\bar{v}}}^\prime \bm{S}\bm{v}
+ \bm{\bar{v}}^\prime \bm{S} \bm{\bar{v}} \right)
& -\frac{1}{NT} \mathbb{E}(\bm{\bar{\bar{v}}}^\prime \bm{S}\bm{\bar{v}})
& 0
\end{bmatrix}
\\
\bm{\gamma_1} &=
\begin{bmatrix}
\frac{1}{NT} \mathbb{E}\!\left(\bm{v}^\prime \bm{S} \bm{v} \right) \\
\frac{1}{NT}\mathbb{E}\!\left(\bm{\bar{v}}^\prime \bm{S} \bm{\bar{v}} \right) \\
\frac{1}{NT}\mathbb{E}\!\left(\bm{\bar{v}}^\prime \bm{S} \bm{v} \right)
\end{bmatrix}.
\end{aligned}
\end{equation*}
Replacing theoretical moment matrix and vector in Equation \eqref{eq:sol2} by the corresponding sample moment matrix and vector yields
\begin{equation} \label{eq:est2}
    \bm{G}_1[\rho_1, \rho_1^2, \sigma_{\bm{\mu}}^2]^{\prime} - \bm{g}_1 = \bm{\xi}_1(\rho_1, \sigma_{\bm{\mu}}^2).
\end{equation}
where expressions for sample moment matrix $\bm{G}_1$ and vector $\bm{g}_1$ are given by
\begin{equation*}
\begin{aligned}
\bm{G_1} &=
\begin{bmatrix}
\frac{2}{NT} \bm{\tilde{\bar{v}}}^\prime \bm{S} \bm{\tilde{v}} 
& -\frac{1}{NT} \bm{\tilde{\bar{v}}}^\prime \bm{S} \bm{\tilde{\bar{v}}}
& 1 \\
\frac{2}{NT} \bm{\tilde{\bar{\bar{v}}}}^\prime \bm{S} \bm{\tilde{\bar{v}}}
& -\frac{1}{NT} \bm{\tilde{\bar{\bar{v}}}}^\prime \bm{S} \bm{\tilde{\bar{\bar{v}}}} 
& \frac{1}{N} \operatorname{tr}(\bm{W}^\prime \bm{W}) \\
\frac{1}{NT} \left(\bm{\tilde{\bar{\bar{v}}}}^\prime \bm{S} \bm{\tilde{v}}
+ \bm{\tilde{\bar{v}}} ^\prime \bm{S} \bm{\tilde{\bar{v}}}  \right)
& -\frac{1}{NT}\bm{\tilde{\bar{\bar{v}}}}^\prime \bm{S} \bm{\tilde{\bar{v}}}
& 0
\end{bmatrix}
\\
\bm{g_1} &=
\begin{bmatrix}
\frac{1}{NT} \bm{\tilde{v}}^\prime \bm{S} \bm{\tilde{v}} \\
\frac{1}{NT} \bm{\tilde{\bar{v}}}^\prime \bm{S} \bm{\tilde{\bar{v}}} \\
\frac{1}{NT} \bm{\tilde{\bar{v}}}^\prime \bm{S} \bm{\tilde{v}}
\end{bmatrix}.
\end{aligned}
\end{equation*}

The GMM estimators can thus be obtained by solving following optimization problems via non-linear least squares estimation
\begin{align}\label{eq:opt}
(\hat{\rho_2}, \hat{\sigma}_{\bm{\epsilon}}^2) &= \argmin_{\rho_2, \sigma_{\bm{\epsilon}}^2} \bm{\xi}_0(\rho_2, \sigma_{\bm{\epsilon}}^2)^{\prime} \bm{\xi}_0(\rho_2, \sigma_{\bm{\epsilon}}^2) \\
(\hat{\rho_1}, \hat{\sigma}_{\bm{\mu}}^2) &= \argmin_{\rho_1, \sigma_{\bm{\mu}}^2} \bm{\xi}_1(\rho_1, \sigma_{\bm{\mu}}^2)^{\prime} \bm{\xi}_1(\rho_1, \sigma_{\bm{\mu}}^2) . 
\end{align}
Let Assumptions \ref{ass:hom} to \ref{ass:away} hold. 
\begin{assumption} \label{ass:away}
    The smallest eigenvalues of $\bm{\Gamma_0}^\prime\bm{\Gamma_0}$ and $\bm{\Gamma_1}^\prime\bm{\Gamma_1}$ are bounded away from zero.
\end{assumption}
Then, the non-linear least squares estimators $\hat{\rho}_1$, $\hat{\rho}_2$, $\hat{\sigma}_{\bm{\mu}}^2$ and $\hat{\sigma}_{\bm{\epsilon}}^2$ are consistent estimators of $\rho_1$, $\rho_2$, $\sigma_{\bm{\mu}}^2$ and $\sigma_{\bm{\epsilon}}^2$ in the sense that $(\hat{\rho_2}, \hat{\sigma}_{\bm{\epsilon}}^2) \to^p (\rho_2, \sigma_{\bm{\epsilon}}^2)$ and $(\hat{\rho_1}, \hat{\sigma}_{\bm{\mu}}^2) \to^p (\rho_1, \sigma_{\bm{\mu}}^2)$ for $N \to \infty$ sufficiently large \citep{kapoor2007, baltagi2016}.

\subsection{Feasible Model-based Gradient Boosting}

\subsubsection{Algorithm and Implementation}
Given estimates of the spatial autocorrelation parameters and the variances, the empirical counterparts $\bm{\hat{\Omega}}$ and $\bm{\hat{\Psi}}$ of $\bm{\Omega}$ and $\bm{\Psi}$ can be computed, enabling a feasible implementation of the model-based gradient boosting algorithm, where the loss function and its corresponding negative gradient determine the updates of the linear predictor. In particular, model-based gradient boosting initializes with an empty model, computes the negative gradient vector of the loss function at the current estimate of the linear predictor and then continues by sequentially fitting specified base-learners to the computed negative gradient vector. In this case, the each base-learner describes the functional form of a single variable occurring in the design matrix, which can be identified as a simple linear model for the GSPECM. For each specified base-learner, the residual sum of squares is computed and the one which minimizes the residual sum of squares is selected. Afterwards, an update in the linear predictor is performed by adding a small fraction of the best-performing base-learner. This algorithm is then continued iteratively until the number of boosting iterations $m_{\text{stop}}$ is reached \citep{friedman2001, bühlmann2007, mayr2014, hepp2016}.

To implement model-based gradient boosting for cross-sectional spatial regression models, a so-called spatial error family can be utilized, which incorporates the concentration matrix into the updates of the algorithm. Although the proposed spatial error family is quite general and can be utilized for model-based gradient boosting for the GSPECM without any alteration, the implementation has various limitations. First, performance of variable selection and regularization is influenced by magnitude of the spatial autocorrelation parameters through the concentration matrix in the updates. Particularly, large positive values of $\rho_1$ and $\rho_2$ may lead to stronger regularization and more pronounced variable selection, which can have a negative impact on the bias-variance trade-off. Second, the concentration matrix severely destabilizes the negative gradient vector in the updates. Although this is rarely a concern in the case of cross-sectional spatial regression models, the increase in complexity in the structure of the concentration matrix for the GSPECM leads to situations, in which the model-based gradient boosting takes very tiny steps towards the solution, thereby requiring a large number of number of boosting iterations. Third, in practice, the implementation via the spatial error family is slow and computationally expensive. Fourth, a formal theoretical justification of model-based gradient boosting with the squared Mahalanobis distance, where the concentration matrix is induced by spatial autocorrelation is still missing \citep{balzer2025}.

To circumvent the discussed limitations, a Cochrane–Orcutt type spatial transformation can be directly applied to the spatial data such that the concentration matrix is not required within the iterations of the model-based gradient boosting algorithm. Consequently, the squared Mahalanobis distance reduces to the $L_2$ loss function and the negative gradient vector to the negative residuals of the Cochrane–Orcutt type spatial transformed variables.\footnote{Essentially, the squared Mahalanobis distance can be interpreted as the $L_2$ loss function weighted by the concentration matrix of the disturbance term. Consequently, the negative gradient corresponds to the weighted negative residuals.} The model-based gradient boosting algorithm thus gains in computational efficiency and formal theoretical justification in low- and high-dimensional settings (see, for example, \cite{zhang2005, bühlmann2006, bühlmann2007, hao2024}).\footnote{Note that such a Cochrane–Orcutt type spatial transformation can also be applied in cross-sectional spatial regression models with autoregressive disturbances, thereby ensuring computational efficiency and theoretical justification of model-based gradient boosting in low- and high-dimensional linear settings.} A summary of the Cochrane–Orcutt type spatial transformed variables as well as the corresponding ingredients for model-based gradient boosting in the random and fixed specification can be seen in Table \ref{tab:transformer}.

\begin{table}[H]
\caption{\label{tab:transformer}
Response variable, design matrix, linear predictor, loss function, negative gradient and concentration matrix utilized in the model-based gradient boosting algorithm for the generalized spatial panel model with error components (GSPECM).}

\resizebox{\textwidth}{!}{%
\begin{minipage}{\textwidth}
\centering
\textbf{Random effects specification} \\[1ex]

\begin{tabular}{lll}
\\[-1.8ex] \hline
\hline \\[-1.8ex]
 & Original & Transformed \\
\hline \\[-1.8ex]
Response variable & $\bm{y}$ & $\bm{y}_{*} = \bm{\hat\Omega}^{-\frac{1}{2}} \bm{y}$ \\
Design matrix     & $\bm{Z}$ & $\bm{Z}_{*} = \bm{\hat\Omega}^{-\frac{1}{2}} \bm{Z}$ \\
Linear predictor  & $\bm{\eta} = \bm{Z}\bm{\delta}$ & $\bm{\eta}_{*} = \bm{Z}_{*}\bm{\delta}$ \\
Loss function     & $f(\bm{y}, \bm{\eta}, \bm{\hat\Omega})  = (\bm{y} - \bm{\eta})^{\prime} \bm{\hat\Omega}^{-1} (\bm{y} - \bm{\eta})$ & $f(\bm{y}_{*}, \bm{\eta}_{*}) =(\bm{y}_{*} - \bm{\eta}_{*})^{\prime}(\bm{y}_{*} - \bm{\eta}_{*})$ \\
Negative gradient & $-\frac{\partial}{\partial \bm{\eta}} f(\bm{y}, \bm{\eta}, \bm{\hat\Omega}) = 2 \bm{\hat\Omega}^{-1} (\bm{y} - \bm{\eta})$ & $-\frac{\partial}{\partial \bm{\eta}_{*}} f(\bm{y}_{*}, \bm{\eta}_{*}) = 2(\bm{y}_{*} - \bm{\eta}_{*})$ \\
Concentration matrix
& \multicolumn{2}{p{0.65\textwidth}}{$\bm{\hat\Omega}^{-1} = \bm{\bar J}_T \otimes \left[T\hat\sigma_{\bm{\mu}}^2 (\bm{\hat{A}}^\prime \bm{\hat{A}})^{-1} + \hat\sigma_{\bm{\epsilon}}^2(\bm{\hat{B}}^\prime \bm{\hat{B}})^{-1}\right]^{-1} + \frac{1}{\hat\sigma_{\bm{\epsilon}}^2} \left[\bm{E}_T \otimes (\bm{\hat{B}}^\prime \bm{\hat{B}})\right]$ with $\bm{\hat{A}} = (\bm{I}_N - \hat\rho_1 \bm{W})$ and $\bm{\hat{B}} = (\bm{I}_N - \hat\rho_2 \bm{W})$}\\
\hline \\[-1.8ex]
\end{tabular}

\vspace{1em}

\textbf{Fixed effects specification} \\[1ex]

\begin{tabular}{lll}
\\[-1.8ex] \hline
\hline \\[-1.8ex]
 & Original & Transformed \\
\hline \\[-1.8ex]
Response variable & $\bm{y}$ & $\bm{y}_{*} = (\bm{E}_T \otimes \bm{\hat{B}}) \bm{y}$ \\
Design matrix     & $\bm{Z}$ & $\bm{Z}_{*} = (\bm{E}_T \otimes \bm{\hat{B}}) \bm{Z}$ \\
Linear predictor  & $\bm{\eta} = \bm{Z}\bm{\delta}$ & $\bm{\eta}_{*} = \bm{Z}_{*}\bm{\delta}$ \\
Loss function     & $f(\bm{y}, \bm{\eta}, \bm{\hat\Psi}) = (\bm{y} - \bm{\eta})^{\prime} \bm{\hat\Psi}^{-1} (\bm{y} - \bm{\eta})$ & $f(\bm{y}_{*}, \bm{\eta}_{*}) = (\bm{y}_{*} - \bm{\eta}_{*})^{\prime}(\bm{y}_{*} - \bm{\eta}_{*})$ \\
Negative gradient & $-\frac{\partial}{\partial \bm{\eta}} f(\bm{y}, \bm{\eta}, \bm{\hat\Psi}) = 2 \bm{\hat\Psi}^{-1} (\bm{y} - \bm{\eta})$ & $-\frac{\partial}{\partial \bm{\eta}_{*}} f(\bm{y}_{*}, \bm{\eta}_{*}) = 2(\bm{y}_{*} - \bm{\eta}_{*})$ \\
Concentration matrix & \multicolumn{2}{p{0.6\textwidth}}{$\bm{\hat\Psi}^{-1} = \frac{1}{\hat\sigma_{\bm{\epsilon}}^2} (\bm{E}_T \otimes (\bm{\hat{B}}^\prime \bm{\hat{B}}))$ with $\bm{\hat{B}} = (\bm{I}_N - \hat\rho_2 \bm{W})$}  \\
\hline \\[-1.8ex]
\end{tabular}

\end{minipage}
}

\vspace{1em}

\begin{minipage}{\textwidth}
\footnotesize
\textbf{Notes:} Entries are reported for the random effects specification and the fixed effects specification with the model on the untransformed scale (Original) and on the Cochrane--Orcutt type spatial transformation scale (Transformed).
\end{minipage}

\end{table}

Applying the Cochrane–Orcutt-type spatial transformation reduces the general model-based gradient boosting algorithm to $L_2$-boosting (LTB) using the $L_2$ loss function, as discussed by \cite{hao2024} in the context of linear panel data models to address the many-instrument problem. The following algorithmic outline shows how LTB can be directly applied for the GSPECM via the Cochrane–Orcutt type transformation:
\begin{enumerate}
  \item Set $m = 0$. Choose offset values for the linear predictor as $\bm{\hat\eta}_{*}^{[0]} = (\bm{0})_{\{i = 1, \dots, NT\}}$ and specify the set of base-learners as linear models  $\delta_1 \bm{Z}_{*1}, \dots, \delta_{K}\bm{Z}_{*K}$, where $K$ is the cardinality of the set of base-learners.
  \item For $m = 1$ to $m_{\text{stop}}$:
  \begin{enumerate}
    \item Calculate the residuals of the $L_2$ loss function at the current estimate of the linear predictor $\bm{\hat\eta}_{*}^{[m - 1]}$
    \begin{equation*}
      \bm{d}^{[m]} = \left(d_{i}^{[m]}\right)_{i = 1, \dots, NT} 
      = \left( -\frac{\partial}{\partial \bm{\eta}_{*}} f(\bm{y}_{*}, \bm{\eta}_{*}) 
      \Bigg|_{\bm{\eta}_{*} = \bm{\hat\eta}_{*}^{[m - 1]}} \right).
    \end{equation*}
    
    \item Fit each specified linear model base-learner to the residuals separately.

    \item Evaluate the residual sum of squares for each base-learner and pick component $j_{*}$ that fits residuals best
    \begin{equation*}
              j_{*} = \operatorname*{argmin}_{1 \leq j \leq K} 
          \sum_{i=1}^{NT} \left(d^{[m]}_i - \hat{\delta}^{m}_{j}\bm{Z}_{*j}\right)^2.
    \end{equation*}

    \item Update the estimate of the linear predictor $\bm{\hat\eta}_{*}^{[m]}$ based on $j_{*}$ via
    \begin{equation*}
    \bm{\hat\eta}_{*}^{[m]} = \bm{\hat\eta}_{*}^{[m - 1]} 
      + s \cdot \hat\delta^{m}_{j^{*}} \bm{Z}_{*j_{*}},
    \end{equation*}
    where $s$ is the learning rate.
  \end{enumerate}
\end{enumerate}

\subsubsection{Optimal Stopping Criterion and Post-hoc Deselection}
In model-based gradient boosting, model and variable selection are controlled by the stopping criterion $m_{\text{stop}}$, which governs the bias-variance trade-off. Stopping too early increases bias, whereas excessive iterations inflate variance and induce overfitting. Arbitrary selection of $m_{\text{stop}}$ compromises both predictive power and the reliability of coefficient estimates, making the identification of an optimal stopping criterion $m_{\text{opt}}$ essential.

In cross-sectional linear models, $m_{\text{opt}}$ is typically determined using information criteria, such as the Akaike information criterion, or resampling techniques, including cross-validation, bootstrapping, or subsampling. For grouped, longitudinal, or panel data, extensions have been proposed in the form of an adjusted Akaike information criterion and an adjusted cross-validation criterion \citep{hao2024, knieper2025}. However, the adjusted criteria rely on assumptions that are violated in the presence of spatial dependence structure in data. As pointed out by \cite{schratz2019}, random cross-validation presumes independence of observations when constructing folds. In spatial data, observations correspond to geographic locations and are typically spatially autocorrelated, such that random fold assignment places nearby observations in both the training and test sets. This violation leads to systematically underestimated prediction errors and unreliable model and variable selection. Furthermore, the adjusted Akaike information criterion is infeasible in this context, as it requires the computation of the degrees of freedom in each boosting iteration, which are currently available only for linear models without spatial dependence.

In contrast, spatial cross-validation techniques provide a coherent alternative for selecting $m_{\text{opt}}$ in the presence of a spatial dependence structure in data. Instead of randomly assigning observations to folds, spatial cross-validation constructs training and test sets that are spatially separated. For example, \cite{brenning2012} generates spatially disjoint folds by clustering spatial polygons based on geodesic distances, while \cite{valavi2019} partition the study region into spatial blocks or buffer zones. In spatial panel models, the temporal dimension can be additionally exploited by leaving out entire time periods, or by jointly excluding spatial and temporal units \citep{schratz2024}. By preventing spatially proximate observations from appearing in both training and test sets, spatial cross-validation avoids downward-biased prediction error estimates, which accurately reflects out-of-sample performance under spatial dependence \citep{balzer2025}.

However, standard and spatial cross-validation techniques often select too many variables, resulting in less parsimonious models \citep{mayr2012}. To obtain more parsimonious models and prevent overfitting, the deselection algorithm of \cite{stromer2022} is applied to LTB for the GSPECM. The algorithm is conducted in two stages. First, model-based gradient boosting is run once and $m_{\text{opt}}$ is selected using spatial cross-validation techniques. Second, the contribution of each base-learner to the cumulative reduction in empirical risk up to $m_{\text{opt}}$ is quantified. Base-learners with negligible contributions are removed and LTB is rerun using only the retained variables while keeping $m_{\text{opt}}$ fixed.

Specifically, let $\mathds{1}(\cdot)$ denote the indicator function, $r^{[m]}$ the empirical risk based on the $L_2$-loss, $(r^{[m-1]} - r^{[m]})$ the risk reduction and $j_{*}^{[m]}$ the base-learner selected in iteration $m$. The attributable risk reduction for base-learner $j$ after $m_{\text{opt}}$ boosting iterations is defined as
\begin{equation}
R_j = \sum_{m=1}^{m_{\text{opt}}}
\mathds{1}\{j = j_{*}^{[m]}\}
\left(r^{[m-1]} - r^{[m]}\right),
\quad j = 1,\dots,K.
\end{equation}
The attributable risk reduction $R_j$ quantifies the portion of the total empirical risk reduction attributable to base-learner $j$ across $m_{\text{opt}}$. The total risk reduction is given by $(r^{[0]} - r^{[m_{\text{opt}}]})$ and a base-learner $j$ is removed if its relative contribution falls below a pre-specified threshold $\tau \in (0,1)$
\begin{equation} \label{eq:des}
R_j < \tau \left(r^{[0]} - r^{[m_{\text{opt}}]}\right).
\end{equation}
Equation \eqref{eq:des} ensures that only base-learners contributing at least a fraction $\tau$ of the total risk reduction are retained in the model. The specific choice of $\tau$ directly controls the sparsity of the final model. Smaller values retain more variables, while larger values enforce more aggressive deselection. For linear and spatial regression models, relative small thresholds $\tau \leq 0.025$ have been found to be appropriate \citep{stromer2022, balzer2025}.

\section{Monte Carlo Experiments} \label{sec:sim}
\subsection{Study Design}
To assess the performance of LTB for the GSPECM, Monte Carlo experiments are conducted. In particular, both the accuracy of parameter estimation and the effectiveness of variable selection and deselection are evaluated in low- and high-dimensional settings. The number of time periods is fixed at $T = 5$, in which the number of observations is set to $N = 100$, resulting in a total of $NT = 500$ observations. The number of independent variables is varied between $K = 40$ and $K = 800$, corresponding to a low- $(NT > K)$ and high-dimensional $(NT < K)$ setting, respectively. Let $\bm{\iota}_{NT}$ be a $NT \times 1$ vector of ones, then the data generating process is given by 
\begin{equation*}
\begin{aligned}
    \bm{y} &= \bm{\iota}_{NT} + 3.5\bm{x}_1 -2.5 \bm{x}_2 -4 (\bm{I}_T \otimes\bm{W}) \bm{x}_1 + 3 (\bm{I}_T \otimes\bm{W})  \bm{x}_2 + \bm{v} \\
    \bm{v} &= \bm{L}_{\bm{\mu}}\bm{u}_1 + \bm{u}_2 \\
    \bm{u}_1 &= \rho_1  \bm{W} \bm{u}_1 + \bm{\mu} \\
    \bm{u}_{2} &= \rho_2  (\bm{I}_T \otimes \bm{W}) \bm{u}_{2} + \bm{\epsilon}.
\end{aligned}
\end{equation*}
Entries of the independent variables $x_{it}$ are generated according to $x_{it} = \zeta_i + \kappa_{it}$ where $\zeta_i \sim \mathcal{U}(-7.5,7.5)$ and $\kappa_{it} \sim \mathcal{U}(-5,5)$ are independently and identically drawn from the uniform distribution. The location-specific random effects and idiosyncratic random innovations are drawn independently and identically from the normal distribution $\mu_i \sim \mathcal{N}(0, \sigma_{\bm{\mu}}^2)$ and $\epsilon_{it} \sim \mathcal{N}(0, \sigma_{\bm{\epsilon}}^2)$ with $\sigma_{\bm{\mu}}^2 = \sigma_{\bm{\epsilon}}^2 = 10$. The spatial weight matrix $\bm{W}$ has a 10-nearest neighbor structure where each location is connected to its ten geographically closest neighbors based on constant centroids of the spatial polygons available in the shapefile, which is based on the $100$ counties in North Carolina. Afterward, $\bm{W}$ is row-normalized such that each row sums up to one\footnote{The shapefile for the North Carolina counties is available in the \texttt{spData} package in R \citep{pebesma2023, bivand2025}.} Spatial autocorrelation parameters vary over a set of $\rho_1, \rho_2 \in \{-0.8,-0.6,-0.4,-0.2,0,0.2,0.4,0.6,0.8\}$. In LTB, the corresponding base-learners are specified as simple linear regression models due to the nature of the data generating process. The learning rate is set to $s = 0.1$, as is standard in the literature (see, for example, \cite{schmid2008, mayr2012, hofner2014}). The optimal stopping criterion $m_{\text{opt}}$ is found by minimizing the empirical risk via 5-fold spatial cross-validation as proposed by \cite{brenning2012}. The threshold parameter in deselection is set to $\tau = 0.01$. Results are reported for ML, GMM, LTB and LTB with post-hoc deselection (DES) for the random effects specification (Random) and fixed effects specification (Fixed) respectively. In each Monte Carlo experiment, a total of $N_{\text{sim}} = 100$ repetitions are conducted.

Regarding the performance of variable selection and deselection, the criteria are chosen based on the confusion matrix. In particular, the reported variable selection criteria are the true positive rate (TPR), which is the proportion of correctly selected variables out of all true informative variables and the true negative rate (TNR), which is the proportion of correctly non-selected variables out of all true non-informative variables \citep{stehmann1997}.

Estimation accuracy is evaluated by reporting the mean squared error (MSE) of $\bm{\delta}$ defined as
\begin{equation*}
    \text{MSE} = \frac{1}{N_{\text{sim}}} \sum_{i = 1}^{N_{\text{sim}}} (\hat{\bm{\delta}}_i - \bm{\delta})^2 
\end{equation*}
For all proposed performance criteria, lower values are always preferred \citep{morris2019}.

Monte Carlo experiments are conducted in the programming language R \citep{R}.
The MLE and GMM estimation of the GSPECM in the low-dimensional setting is performed via the \texttt{splm} package \citep{millo2012, bivand2021}. Model-based gradient boosting for generalized, additive and interaction models can be found in the \texttt{mboost} package \citep{bühlmann2007, hothorn2010, hofner2014, hofner2015}. Spatial cross-validation is performed via the \texttt{sperrorest} package \citep{brenning2012}. An implementation for LTB for the GSPECM incorporating the deselection algorithm and the R code for reproducibility of all Monte Carlo experiments can be found in the GitHub repository \url{https://github.com/micbalz/SpatPanelRegBoost}.

\subsection{Results}
\subsubsection{Low-dimensional Linear Setting}
In Table \ref{tab:res_low}, the results for the performance of estimation and variable selection in the low-dimensional linear settings over $N_{\text{sim}} = 100$ repetitions are presented. Across all combinations of the spatial autocorrelation parameters $(\rho_1,\rho_2)$, TPR equals one for all methods in both random and fixed effects specifications, indicating that all informative variables are selected in this low-dimensional linear setting. Differences between methods arise exclusively from their ability to correctly exclude non-informative variables, as reflected by the TNR and from their estimation accuracy as measured by the MSE.

ML and GMM estimators exhibit a TNR of zero throughout, which is unsurprising since these methods do not encompass an inherent variable selection mechanism. In contrast, LTB substantially improves variable selection performance, achieving moderate to high TNR values that generally increase with the strength of spatial dependence. DES further improves on sparsity, yielding perfect exclusion of non-informative variables across all $(\rho_1,\rho_2)$ combinations in both random and fixed effects specifications.

These improvements in variable selection translate directly into gains in estimation accuracy. While ML and GMM display comparatively large MSE values, which increase with stronger spatial autocorrelation, both LTB and DES lead to considerable reductions in MSE values. Among them, DES achieves the lowest MSE values across all $(\rho_1,\rho_2)$ combinations. The fixed effects specification generally exhibits larger MSE values than the corresponding random effects specification.

\begin{table}[H]
\caption{\label{tab:res_low}
Performance of estimation and variable selection in the low-dimensional linear setting ($N = 100$, $T = 5$, $NT = 500$, $K = 40$) for a set of spatial autocorrelation parameters $(\rho_1,\rho_2)$ with $N_{\text{sim}} = 100$ repetitions.}
\centering
\setlength{\tabcolsep}{6pt}
\begin{tabular}{ccclccccccc}
\\[-1.8ex] \hline
\hline \\[-1.8ex]
& & & & \multicolumn{3}{c}{Random} & \multicolumn{3}{c}{Fixed} \\
\cmidrule(lr){5-7} \cmidrule(lr){8-10}
$\rho_1$ & $\rho_2$ & & ML & GMM & LTB & DES & GMM & LTB & DES \\
\hline \\[-1.8ex]

\multirow{3}{*}{$-0.2$} & \multirow{3}{*}{$0.2$}
& TPR & 1.000 & 1.000 & 1.000 & 1.000 & 1.000 & 1.000 & 1.000 \\
& & TNR & 0.000 & 0.000 & 0.718 & 1.000 & 0.000 & 0.764 & 1.000 \\
& & MSE & 0.321 & 0.338 & 0.108 & 0.043 & 0.515 & 0.135 & 0.055 \\
\hline

\multirow{3}{*}{$-0.4$} & \multirow{3}{*}{$0.4$}
& TPR & 1.000 & 1.000 & 1.000 & 1.000 & 1.000 & 1.000 & 1.000 \\
& & TNR & 0.000 & 0.000 & 0.743 & 1.000 & 0.000 & 0.791 & 1.000 \\
& & MSE & 0.355 & 0.374 & 0.107 & 0.045 & 0.563 & 0.142 & 0.062 \\
\hline

\multirow{3}{*}{$-0.6$} & \multirow{3}{*}{$0.6$}
& TPR & 1.000 & 1.000 & 1.000 & 1.000 & 1.000 & 1.000 & 1.000 \\
& & TNR & 0.000 & 0.000 & 0.761 & 1.000 & 0.000 & 0.803 & 1.000 \\
& & MSE & 0.427 & 0.474 & 0.123 & 0.052 & 0.705 & 0.148 & 0.074 \\
\hline

\multirow{3}{*}{$-0.8$} & \multirow{3}{*}{$0.8$}
& TPR & 1.000 & 1.000 & 1.000 & 1.000 & 1.000 & 1.000 & 1.000 \\
& & TNR & 0.000 & 0.000 & 0.808 & 1.000 & 0.000 & 0.824 & 1.000 \\
& & MSE & 0.503 & 0.606 & 0.125 & 0.064 & 0.915 & 0.158 & 0.086 \\
\hline

\multirow{3}{*}{$0$} & \multirow{3}{*}{$0$}
& TPR & 1.000 & 1.000 & 1.000 & 1.000 & 1.000 & 1.000 & 1.000 \\
& & TNR & 0.000 & 0.000 & 0.667 & 1.000 & 0.000 & 0.758 & 1.000 \\
& & MSE & 0.308 & 0.341 & 0.125 & 0.038 & 0.429 & 0.136 & 0.059 \\
\hline

\multirow{3}{*}{$0.2$} & \multirow{3}{*}{$-0.2$}
& TPR & 1.000 & 1.000 & 1.000 & 1.000 & 1.000 & 1.000 & 1.000 \\
& & TNR & 0.000 & 0.000 & 0.689 & 1.000 & 0.000 & 0.742 & 1.000 \\
& & MSE & 0.268 & 0.295 & 0.100 & 0.034 & 0.395 & 0.122 & 0.047 \\
\hline

\multirow{3}{*}{$0.4$} & \multirow{3}{*}{$-0.4$}
& TPR & 1.000 & 1.000 & 1.000 & 1.000 & 1.000 & 1.000 & 1.000 \\
& & TNR & 0.000 & 0.000 & 0.714 & 1.000 & 0.000 & 0.738 & 1.000 \\
& & MSE & 0.270 & 0.319 & 0.115 & 0.042 & 0.371 & 0.133 & 0.052 \\
\hline

\multirow{3}{*}{$0.6$} & \multirow{3}{*}{$-0.6$}
& TPR & 1.000 & 1.000 & 1.000 & 1.000 & 1.000 & 1.000 & 1.000 \\
& & TNR & 0.000 & 0.000 & 0.723 & 1.000 & 0.000 & 0.725 & 1.000 \\
& & MSE & 0.237 & 0.311 & 0.101 & 0.039 & 0.383 & 0.122 & 0.039 \\
\hline

\multirow{3}{*}{$0.8$} & \multirow{3}{*}{$-0.8$}
& TPR & 1.000 & 1.000 & 1.000 & 1.000 & 1.000 & 1.000 & 1.000 \\
& & TNR & 0.000 & 0.000 & 0.757 & 1.000 & 0.000 & 0.733 & 1.000 \\
& & MSE & 0.243 & 0.378 & 0.127 & 0.049 & 0.416 & 0.137 & 0.054 \\
\hline \\[-1.8ex]
\end{tabular}

\begin{minipage}{\textwidth}
\footnotesize
\textbf{Notes:} Reported are the root mean squared error of $\bm{\delta}$ (MSE), true positive rate (TPR), and true negative rate (TNR) for maximum likelihood (ML), generalized method of moments (GMM), $L_2$-boosting (LTB) and $L_2$-boosting with post-hoc deselection (DES) for the random effects specification (Random) and fixed effects specification (Fixed).
\end{minipage}
\end{table}

\subsubsection{High-dimensional Linear Setting}
In Table \ref{tab:res_high}, the results for the performance of estimation and variable selection in the high-dimensional linear settings over $N_{\text{sim}} = 100$ repetitions are presented. Similar to the low-dimensional linear settings, both LTB and DES achieve a TPR equal to one under random and fixed effects specifications across all combinations of the spatial autocorrelation parameters $(\rho_1,\rho_2)$. This indicates that all informative variables are successfully identified even in the high-dimensional linear setting, highlighting proper functionality of LTB and DES.

Differences between LTB and DES are primarily reflected in their ability to exclude non-informative variables and in their estimation accuracy. ML and GMM estimators are not applicable in this setting and are therefore omitted. For LTB, the TNR is high, ranging between approximately $0.96$ and $0.98$, and shows a mild increase with the strength of spatial dependence. DES further improves variable selection performance, yielding perfect exclusion of non-informative variables across all $(\rho_1,\rho_2)$ combinations in both random and fixed effect specifications.

\begin{table}[H]
\caption{\label{tab:res_high}
Performance of estimation and variable selection in the high-dimensional linear setting ($N = 100$, $T = 5$, $NT = 500$, $K = 800$) for a set of spatial autocorrelation parameters $(\rho_1,\rho_2)$ with $N_{\text{sim}} = 100$ repetitions.}
\centering
\setlength{\tabcolsep}{6pt}
\begin{tabular}{ccclccccccc}
\\[-1.8ex] \hline
\hline \\[-1.8ex]
& & & & \multicolumn{3}{c}{Random} & \multicolumn{3}{c}{Fixed} \\
\cmidrule(lr){5-7} \cmidrule(lr){8-10}
$\rho_1$ & $\rho_2$ & & ML & GMM & LTB & DES & GMM & LTB & DES \\
\hline \\[-1.8ex]

\multirow{3}{*}{$-0.2$} & \multirow{3}{*}{$0.2$}
& TPR & -- & -- & 1.000 & 1.000 & -- & 1.000 & 1.000 \\
& & TNR & -- & -- & 0.971 & 1.000 & -- & 0.968 & 1.000 \\
& & MSE & -- & -- & 0.190 & 0.056 & -- & 0.262 & 0.067 \\
\hline

\multirow{3}{*}{$-0.4$} & \multirow{3}{*}{$0.4$}
& TPR & -- & -- & 1.000 & 1.000 & -- & 1.000 & 1.000 \\
& & TNR & -- & -- & 0.974 & 1.000 & -- & 0.970 & 1.000 \\
& & MSE & -- & -- & 0.207 & 0.069 & -- & 0.285 & 0.081 \\
\hline

\multirow{3}{*}{$-0.6$} & \multirow{3}{*}{$0.6$}
& TPR & -- & -- & 1.000 & 1.000 & -- & 1.000 & 1.000 \\
& & TNR & -- & -- & 0.979 & 1.000 & -- & 0.974 & 1.000 \\
& & MSE & -- & -- & 0.182 & 0.066 & -- & 0.240 & 0.084 \\
\hline

\multirow{3}{*}{$-0.8$} & \multirow{3}{*}{$0.8$}
& TPR & -- & -- & 1.000 & 1.000 & -- & 1.000 & 1.000 \\
& & TNR & -- & -- & 0.980 & 1.000 & -- & 0.978 & 1.000 \\
& & MSE & -- & -- & 0.183 & 0.072 & -- & 0.252 & 0.106 \\
\hline

\multirow{3}{*}{$0$} & \multirow{3}{*}{$0$}
& TPR & -- & -- & 1.000 & 1.000 & -- & 1.000 & 1.000 \\
& & TNR & -- & -- & 0.974 & 1.000 & -- & 0.970 & 1.000 \\
& & MSE & -- & -- & 0.183 & 0.056 & -- & 0.263 & 0.064 \\
\hline

\multirow{3}{*}{$0.2$} & \multirow{3}{*}{$-0.2$}
& TPR & -- & -- & 1.000 & 1.000 & -- & 1.000 & 1.000 \\
& & TNR & -- & -- & 0.971 & 1.000 & -- & 0.965 & 1.000 \\
& & MSE & -- & -- & 0.193 & 0.056 & -- & 0.284 & 0.063 \\
\hline

\multirow{3}{*}{$0.4$} & \multirow{3}{*}{$-0.4$}
& TPR & -- & -- & 1.000 & 1.000 & -- & 1.000 & 1.000 \\
& & TNR & -- & -- & 0.971 & 1.000 & -- & 0.962 & 1.000 \\
& & MSE & -- & -- & 0.205 & 0.060 & -- & 0.335 & 0.071 \\
\hline

\multirow{3}{*}{$0.6$} & \multirow{3}{*}{$-0.6$}
& TPR & -- & -- & 1.000 & 1.000 & -- & 1.000 & 1.000 \\
& & TNR & -- & -- & 0.972 & 1.000 & -- & 0.961 & 1.000 \\
& & MSE & -- & -- & 0.214 & 0.069 & -- & 0.358 & 0.081 \\
\hline

\multirow{3}{*}{$0.8$} & \multirow{3}{*}{$-0.8$}
& TPR & -- & -- & 1.000 & 1.000 & -- & 1.000 & 1.000 \\
& & TNR & -- & -- & 0.974 & 1.000 & -- & 0.960 & 1.000 \\
& & MSE & -- & -- & 0.209 & 0.069 & -- & 0.375 & 0.087 \\
\hline \\[-1.8ex]
\end{tabular}

\begin{minipage}{\textwidth}
\footnotesize
\textbf{Notes:} Reported are the root mean squared error of $\bm{\delta}$ (MSE), true positive rate (TPR), and true negative rate (TNR) for maximum likelihood (ML), generalized method of moments (GMM), $L_2$-boosting (LTB) and $L_2$-boosting with post-hoc deselection (DES) for the random effects specification (Random) and fixed effects specification (Fixed).
\end{minipage}
\end{table}

Gains in variable selection performance come with corresponding improvements in estimation accuracy. Although LTB attains moderate MSE values, DES achieves lowest MSE values across all $(\rho_1, \rho_2)$ combinations. Similar to the the low-dimensional linear setting, MSE values are generally higher under the fixed effects specification than under the random effects specification.

Overall, the results demonstrate that in both low and high-dimensional linear settings, LTB combined with DES effectively balances perfect detection of informative variables with accurate exclusion of non-informative variables, resulting in substantial efficiency gains in terms of MSE over classical ML and GMM estimators under both random and fixed effects specifications.

\section{Empirical Illustrations} \label{sec:illu}

\subsection{Non-life Insurance in Italian Provinces}
The data set on non-life insurance consumption across Italian provinces was first examined by \cite{millo2011} and later revisited by \cite{millo2014} and \cite{millo2022}. It focuses on identifying the determinants of per capita equilibrium consumption of non-life insurance across $N_{\text{Italy}} = 103$ Italian provinces over the period $1998-2002$, corresponding to $T_{\text{Italy}} = 5$ years. The data set contains a broad set of province-level socioeconomic characteristics, summarized in Table \ref{tab:var_italy}. Since most of the variables are either time-invariant or exhibit only limited variation over time, a fixed effects specification is not well suited.

\begin{table}[H]
\caption{Data type and description of dependent and independent variables for non-life insurance consumption across Italian provinces data set (see, \cite{millo2011, millo2012, millo2014, millo2022}).}
\label{tab:var_italy}
\centering
\begin{tabular}{@{\extracolsep{5pt}}llp{9.3cm}}
\\[-1.8ex] \hline
\hline \\[-1.8ex]
\textbf{Variable} & \textbf{Type} & \textbf{Description} \\ 
\hline \\[-1.8ex]
PPCD   & Numeric & Real per capita premiums in $2000$ euros, non-life insurance excluding mandatory motor third-party liability \\
AGEN   & Numeric & Density of insurance agencies per $1000$ inhabitants \\
BANK   & Numeric & Real bank deposits per capita \\
DEN    & Numeric & Density of inhabitants per square kilometre \\
FAM    & Numeric & Average number of family members \\
INEF   & Numeric & Judicial inefficiency: average years to settle first degree of civil case \\
RGDP   & Numeric & Real gross domestic product per capita \\
RIRS   & Numeric & Real interest rate on borrowing \\
SCHOOL & Numeric & Share of people with second-grade schooling or more \\
TRUST  & Numeric & Survey results to the question "do you trust others?" \\
VAAGR  & Numeric & Share of value added, agricultural sector \\
\hline \\[-1.8ex]
\end{tabular}
\end{table}

Thus, to investigate the determinants of per capita equilibrium consumption of non-life insurance from a data-driven variable selection perspective via LTB and DES, a variety of random effects models, namely ANS, KKP and GSPECM are considered. For additional comparison, results for classical GMM estimation are reported. All independent variables presented in Table \ref{tab:var_italy} and corresponding spatial lags of independent variables as well as an intercept are included in the estimation, leading to a low-dimensional linear setting with $21$ coefficients to estimate in a non-parsimonious model. Furthermore, continuos variables are transformed via centering and scaling to ensure comparability across variables in terms of the scale and proper convergence of LTB. The setup for LTB and DES follows closely the Monte Carlo experiments in Section \ref{sec:sim}, such that the learning rate is set to $s = 0.1$ and the threshold to $\tau  = 0.01$. The optimal stopping criterion $m_{\text{opt}}$ is found by minimizing the empirical risk via 10-fold spatial cross-validation as discussed by \cite{brenning2012}.\footnote{The shapefile of the Italian provinces is provided freely by the Italian National Institute of Statistics and can be found at \url{https://www.istat.it/notizia/confini-delle-unita-amministrative-a-fini-statistici-al-1-gennaio-2018-2/}.}

\begin{table}[H]
\caption{Coefficient estimates across different estimation strategies for the Italian non-life insurance data set.}
\centering
\label{tab:italy}
\setlength{\tabcolsep}{2.5pt}
\begin{tabular}{lccccccccc}
\\[-1.8ex]\hline
\hline \\[-1.8ex]
 & \multicolumn{3}{c}{GMM} & \multicolumn{3}{c}{LTB} & \multicolumn{3}{c}{DES} \\
\cmidrule{2-4}\cmidrule{5-7}\cmidrule{8-10}
 
& ANS & KKP & GSPECM 
& ANS & KKP & GSPECM
& ANS & KKP & GSPECM \\
\hline \\[-1.8ex]
$\hat\rho_1$ & 0.000 & 0.277 & 0.337 & 0.000 & 0.277 & 0.337 & 0.000 & 0.277 & 0.337 \\
$\hat\rho_2$ & 0.277 & 0.277 & 0.277 & 0.277 & 0.277 & 0.277 & 0.277 & 0.277 & 0.277 \\
$\hat\sigma_{\bm{\mu}}$ & 0.080 & 0.088 & 0.079 & 0.080 & 0.088 & 0.079 & 0.080 & 0.088 & 0.079 \\
$\hat\sigma_{\bm{\epsilon}}$ & 0.019 & 0.019 & 0.019 & 0.019 & 0.019 & 0.019 & 0.019 & 0.019 & 0.019 \\
(Intercept) & -0.004 & -0.001 & -0.005 & -- & -- & -- & -- & -- & -- \\
AGEN        & 0.039 & 0.043 & 0.037 & 0.002 & -- & 0.002 & -- & -- & -- \\
BANK        & 0.132 & 0.125 & 0.132 & 0.112 & 0.104 & 0.112 & 0.130 & 0.114 & 0.129 \\
DEN         & 0.136 & 0.136 & 0.137 & 0.112 & 0.098 & 0.114 & 0.115 & 0.112 & 0.120 \\
FAM         & -0.097 & -0.118 & -0.092 & -0.080 & -0.094 & -0.076 & -0.100 & -0.116 & -0.099 \\
INEF        & -0.125 & -0.119 & -0.125 & -0.103 & -0.097 & -0.106 & -0.131 & -0.112 & -0.131 \\
RGDP        & 0.381 & 0.398 & 0.375 & 0.485 & 0.526 & 0.473 & 0.485 & 0.526 & 0.473 \\
RIRS        & -0.031 & -0.030 & -0.031 & -0.021 & -0.016 & -0.021 & -0.030 & -- & -0.032 \\
SCHOOL      & 0.052 & 0.039 & 0.055 & 0.016 & -- & 0.020 & -- & -- & -- \\
TRUST       & 0.079 & 0.102 & 0.073 & 0.053 & 0.069 & 0.049 & -- & 0.080 & -- \\
VAAGR       & -0.017 & -0.019 & -0.016 & -0.002 & -0.003 & -0.002 & -- & -- & -- \\
$\bm{W}$AGEN   & 0.021 & 0.020 & 0.021 & 0.002 & -- & 0.003 & -- & -- & -- \\
$\bm{W}$BANK   & 0.013 & 0.009 & 0.014 & -- & -- & -- & -- & -- & -- \\
$\bm{W}$DEN    & 0.002 & 0.003 & 0.002 & -- & -- & -- & -- & -- & -- \\
$\bm{W}$FAM    & 0.042 & 0.037 & 0.043 & -- & -- & 0.004 & -- & -- & -- \\
$\bm{W}$INEF   & 0.007 & 0.009 & 0.007 & -- & -- & -- & -- & -- & -- \\
$\bm{W}$RGDP   & 0.004 & 0.013 & 0.003 & 0.017 & 0.018 & 0.016 & -- & -- & -- \\
$\bm{W}$RIRS   & -0.033 & -0.032 & -0.033 & -0.007 & -0.005 & -0.008 & -- & -- & -- \\
$\bm{W}$SCHOOL & 0.036 & 0.032 & 0.037 & 0.020 & 0.016 & 0.021 & -- & -- & -- \\
$\bm{W}$TRUST  & 0.014 & 0.014 & 0.014 & 0.003 & 0.002 & 0.004 & -- & -- & -- \\
$\bm{W}$VAAGR  & 0.013 & 0.015 & 0.012 & -- & -- & -- & -- & -- & -- \\
\hline \\[-1.8ex]
\end{tabular}
\begin{minipage}{\textwidth}
\footnotesize
\textbf{Notes:} Reported are the results for the generalized method of moments (GMM), $L_2$-boosting (LTB) and LTB with post-hoc deselection (DES) for \cite{anselin1988} random effects model (ANS), \cite{kapoor2007} random effects model (KKP) and the generalized spatial panel model with error components (GSPECM).
\end{minipage}
\end{table}

The estimated coefficients for all estimation strategies and different model specifications can be seen in Table \ref{tab:italy}. Regarding the variable selection performance, the results for all considered strategies are comparable. Particularly, LTB reduces the number of variables from $21$ to $12 - 16$ depending on the particular random effects model. The sparsest results are obtained for the KKP random effects model, while the results for the GSPECM are worst. Further improvement in variable selection can be seen in DES. Particularly, the number of included variables is further reduced to only $6$ across all considered random effects models, where spatial lags of independent variables are generally not selected.

Estimates of the spatial autocorrelation parameters $\hat\rho_1$ and $\hat\rho_2$ are both positive for the GSPECM and KKP random effects model, indicating persistent spatial dependence between disturbances of neighbouring provinces in Italy. Furthermore, estimates of the variances of the disturbances $\hat\sigma_{\bm{\mu}}$ and $\hat\sigma_{\bm{\epsilon}}$ remain comparable across all considered random effects models in terms of magnitude. 

Moreover, the sign and overall direction of the coefficients are similar across all estimation strategies. The main difference lies in their magnitude, which is affected by the shrinkage induced through early stopping in both LTB and DES. Taking the results for DES for the GSPECM as an example, the most important variables explaining the non-life insurance in Italian provinces in the years $1998-2002$ are the the real bank deposits per capita, density of inhabitants per square kilometre, the average number of family members, the judicial inefficiency, the real gross domestic product per capita and the real interest rate on borrowing. Since the independent variables are transformed by simple scaling and centering, the coefficients have an intuitive and simple interpretation. For instance, an increase in average number of family members by one person ceteris paribus decreases the real per capita non-life insurance consumption by $0.099$ euros on average. Consequently, holding all other variables constant, an increase in the gross domestic product per capita by one unit is associated with an average $0.473$ euros increase in real per capita non-life insurance consumption.

\subsection{Rice Production in Indonesian Farms}
As a second empirical illustration, a spatial panel data set of $171$ rice farms in the Cimanuk River Basin, West Java, Indonesia, previously analysed by  \cite{horrace1996, horrace2000, druska2004, millo2011} and \cite{millo2022} is considered. The farms were observed over six growing seasons, three wet and three dry, resulting in a balanced panel with $N_{\text{Indonesia}} = 171$ farms and $T_{\text{Indonesia}} = 6$ seasons. Data were collected by the Agro Economic Survey as part of the Rural Dynamic Study and obtained from the Center for Agro Economic Research, Ministry of Agriculture, Indonesia. The data set includes a range of farm-level socioeconomic, agronomic, and seasonal environmental variables, which are summarized in Table \ref{tab:var_idonesia}. Most considered variable are time-varying, thereby enabling a fixed effects specification.

\begin{table}[H]
\caption{Data type and description of dependent and independent variables for the rice production in Indonesian farms data set (see, for example, \cite{horrace1996, horrace2000, druska2004, millo2012}).}
\label{tab:var_idonesia}
\centering
\begin{tabular}{@{\extracolsep{5pt}}llp{7cm}}
\\[-1.8ex] \hline
\hline \\[-1.8ex]
\textbf{Variable} & \textbf{Type} & \textbf{Description} \\ 
\hline \\[-1.8ex]
GOUTPUT    & Numeric & Gross output of rice in kilogram \\
BIMAS      & Categorical & Part of the Bimbingan Masal intensification program (NO/YES/MIXED) \\
FAMLABOR   & Numeric & Family labor in hours \\
HIREDLABOR & Numeric & Hired labor in hours \\
PESTICIDE  & Numeric & Pesticide cost in Indonesian Rupiah \\
PHOSPHATE  & Numeric & Phosphate in kilogram \\
PPHOSPH    & Numeric & Price of phosphate in Indonesian Rupiah per kilogram \\
PRICE      & Numeric & Price of rough rice in Indonesian Rupiah per kilogram \\
PSEED      & Numeric & Price of seed in Indonesian Rupiah per kilogram \\
PUREA      & Numeric & Price of urea in Indonesian Rupiah per kilogram \\
SEED       & Numeric & Seed in kilogram \\
SIZE       & Numeric & Total area cultivated with rice in hectares \\
STATUS     & Categorical & Land ownership status (OWNER/SHARE/MIXED) \\
TOTLABOR   & Numeric & Total labor hours, excluding harvest labor hours \\
UREA       & Numeric & Urea in kilogram \\
VARIETIES  & Categorical & Varieties of planted rice (TRAD/HIGH/MIXED) \\
WAGE       & Numeric & Labour wage in Indonesian Rupiah per hour \\
\hline \\[-1.8ex]
\end{tabular}
\end{table}

The determinants of rice production on Indonesian farms are investigated using a data-driven variable selection approach based on LTB and DES. Both fixed and random effects specifications of the GSPECM are reported. As a benchmark, GMM estimates for each specification are additionally reported to enable direct comparison. All independent variables listed in Table \ref{tab:var_idonesia}, together with their corresponding spatial lags and an intercept, are included in the estimation. This results in a low-dimensional linear setting with $39$ coefficients to be estimated in a non-parsimonious model. Furthermore, continuous variables are log-transformed or scaled via division by $100$ to ensure comparability across variables in terms of magnitude and to ensure proper convergence of the LTB algorithm. The implementation of LTB and DES follows closely the Monte Carlo experiments in Section \ref{sec:sim}. The learning rate is set to $s = 0.1$ and the threshold parameter is fixed at $\tau  = 0.01$. The optimal stopping criterion $m_{\text{opt}}$ is selected by minimizing the empirical risk using $6$-fold leave-time-out cross-validation as discussed by \cite{schratz2024}.

\begin{table}[!htpb]
\caption{Coefficient estimates across different estimation strategies for the rice production in Indonesian farms data set.}
\centering
\label{tab:indonesia}
\begin{tabular}{lcccccc}
\\[-1.8ex]\hline
\hline \\[-1.8ex]
 & \multicolumn{3}{c}{Random} & \multicolumn{3}{c}{Fixed} \\
\cmidrule{2-4}\cmidrule{5-7}
Variable 
& GMM & LTB & DES 
& GMM & LTB & DES \\
\hline \\[-1.8ex]
$\hat\rho_1$ & 0.989 & 0.989 & 0.989 & -- & -- & -- \\
$\hat\rho_2$ & 0.470 & 0.470 & 0.470 & 0.470 & 0.470 & 0.470 \\
$\hat\sigma_{\bm{\mu}}$ & 0.012 & 0.012 & 0.012 & -- & -- & -- \\
$\hat\sigma_{\bm{\epsilon}}$ & 0.073 & 0.073 & 0.073 & 0.073 & 0.073 & 0.073 \\
(Intercept) & -5.407 & 0.331 & -- & -- & -- & -- \\
BIMASMIXED & -0.082 & -0.038 & -- & -0.076 & -0.065 & -- \\
BIMASYES & 0.086 & 0.029 & -- & 0.050 & 0.023 & -- \\
FAMLABOR & -0.001 & -- & -- & -0.003 & -- & -- \\
HIREDLABOR & 0.011 & 0.012 & -- & 0.010 & 0.017 & -- \\
PESTICIDE & 0.005 & 0.003 & -- & 0.003 & 0.003 & -- \\
PHOSPHATE & 0.716 & 0.600 & -- & 0.693 & 0.768 & -- \\
PPHOSPH & -0.772 & -- & -- & -0.364 & -0.046 & -- \\
PRICE & 0.014 & -- & -- & 0.041 & -- & -- \\
PSEED & -0.057 & -- & -- & -0.044 & -0.039 & -- \\
PUREA & 0.726 & -- & -- & 0.277 & -- & -- \\
SEED & 0.106 & 0.109 & 0.122 & 0.094 & 0.092 & 0.108 \\
SIZE & 0.503 & 0.492 & 0.509 & 0.483 & 0.488 & 0.499 \\
STATUSMIXED & 0.007 & 0.005 & -- & 0.028 & 0.039 & -- \\
STATUSSHARE & 0.155 & -- & -- & 0.200 & 0.153 & -- \\
TOTLABOR & 0.215 & 0.224 & 0.227 & 0.217 & 0.202 & 0.231 \\
UREA & 0.131 & 0.123 & 0.138 & 0.103 & 0.098 & 0.101 \\
VARIETIESHIGH & 0.084 & 0.044 & -- & 0.066 & 0.074 & -- \\
VARIETIESMIXED & 0.117 & 0.026 & -- & 0.128 & 0.126 & -- \\
WAGE & 0.080 & -- & -- & 0.136 & 0.108 & -- \\
$\bm{W}$BIMASMIXED & 0.083 & 0.149 & -- & 0.094 & 0.316 & -- \\
$\bm{W}$BIMASYES & 1.125 & -- & -- & 1.035 & 0.259 & -- \\
$\bm{W}$FAMLABOR & -0.178 & -- & -- & -0.118 & -- & -- \\
$\bm{W}$HIREDLABOR & -0.297 & -0.013 & -- & -0.264 & -0.118 & -- \\
$\bm{W}$PESTICIDE & -0.009 & -- & -- & -0.010 & -0.014 & -- \\
$\bm{W}$PHOSPHATE & 6.329 & 3.942 & -- & 6.848 & 8.977 & 5.473 \\
$\bm{W}$PPHOSPH & -9.808 & 0.597 & 1.230 & -9.153 & -- & -- \\
$\bm{W}$PRICE & -1.017 & -- & -- & -1.044 & -0.570 & -- \\
$\bm{W}$PSEED & 0.669 & -- & -- & 0.617 & 0.148 & -- \\
$\bm{W}$PUREA & 12.717 & 0.306 & -- & 12.097 & 1.938 & 0.901 \\
$\bm{W}$SEED & -0.154 & -- & -- & -0.115 & -- & -- \\
$\bm{W}$SIZE & 0.062 & -- & -- & 0.049 & -0.036 & -- \\
$\bm{W}$STATUSMIXED & -0.472 & 0.241 & -- & -0.510 & -0.300 & -- \\
$\bm{W}$STATUSSHARE & 2.828 & -- & -- & 2.714 & 1.887 & -- \\
$\bm{W}$TOTLABOR & 0.384 & -- & -- & 0.274 & -- & -- \\
$\bm{W}$UREA & 0.133 & -- & -- & 0.144 & -0.036 & -- \\
$\bm{W}$VARIETIESHIGH & -0.688 & -- & -- & -0.633 & -0.235 & -- \\
$\bm{W}$VARIETIESMIXED & 0.387 & -- & -- & 0.433 & 0.640 & -- \\
$\bm{W}$WAGE & -0.174 & -- & -- & -0.210 & -- & -- \\
\hline \\[-1.8ex]
\end{tabular}
\begin{minipage}{\textwidth}
\footnotesize
\textbf{Notes:} Reported are the results for the generalized method of moments (GMM), $L_2$-boosting (LTB) and LTB with post-hoc deselection (DES) for the random effects specification (Random) and fixed effects specification (Fixed) in the generalized spatial panel model with error components (GSPECM).
\end{minipage}
\end{table}

Table \ref{tab:indonesia} presents coefficient estimates for rice production in Indonesian farms under both random and fixed effects specifications of the GSPECM. In terms of variable selection, the results of the considered strategies show some divergence. Specifically, under the random effects specification, LTB reduces the number of variables from $39$ to $19$, whereas under the fixed effects specification, it reduces the number of variables only to $29$, indicating considerably weaker performance. In contrast, the improvement of DES for both random and fixed effects specification are comparable, reducing the number of variables to $5$ and $6$ respectively, including spatial lags of independent variables. Furthermore, the remaining variables are almost identical across both specifications.

Estimates of the spatial autocorrelation parameters $\hat\rho_1$ and $\hat\rho_2$ are both positive in the random effects specification, indicating strong persistent spatial dependence between neighbouring farms in Indonesia. However, the estimate of $\hat{\rho}_1$ lies at the boundary of the admissible parameter range, casting doubt on the inclusion and estimation of the spatial autocorrelation parameter in the location-specific disturbances.

Moreover, the sign and overall direction of the coefficients remain similar across both specifications. The primary difference lies in the magnitude of the coefficients, which varies due to the shrinkage effects introduced by early stopping in LTB and DES. Taking the results for DES as an example, the most important variables explaining the rice production in Indonesian farms are the seed in kilogram, the total area cultivated with rice in hectares, the total labor hours, urea in kilogram, and the price of phosphate in Indonesian Rupiah per kilogram in neighbouring farms. The interpretation is now based on percentages due to the log-transformation. For instance, an one percent increase in seed usage in kilogram ceteris paribus increases rice production by $0.122$ percent in kilogram on average. Consequently, holding all other variables constant, an increase in the cultivated area in hectares by one percent is associated with an average $0.509$ percent increase in rice production in kilogram.

\subsection{Life Expectancy in German Districts}
The final empirical illustration is concerned with investigating the drivers of life expectancy in German districts, which is based on the "Indicators and Maps on Spatial and Urban Development in Germany and Europe" (INKAR) database. The INKAR database contains a variety of socio-demographic, socio-economic, and environmental indicators available on German district level.\footnote{The INKAR database in the current as well as the 2021 version with the codebook for all indicators is freely available at \url{https://www.inkar.de/}} The constructed spatial panel data set includes $N_{\text{Germany}} = 400$ districts in which the average life expectancy of a newborn in years has been observed over a time period of $T_{\text{Germany}} = 5$ years, particularly from $2014-2019$. More than $23$ independent variable of interest can be identified, which are summarized in Table \ref{tab:var_ger}. Most considered variable are time-invariant or have little variability over time, deeming a fixed effects specification inappropriate \citep{BBSR2024}.

\begin{table}[!htpb]
\caption{Data type and description of dependent and independent variables for German district data set (see, \cite{balzer2025}).}
\label{tab:var_ger}
\begin{tabular}{@{\extracolsep{5pt}}llp{7.5cm}}
\\[-1.8ex] \hline
\hline \\[-1.8ex]
\textbf{Variable} & \textbf{Type} & \textbf{Description} \\ 
\hline \\[-1.8ex]
LIFE       & Numeric &  Average life expectancy of a newborn in years\\
ACADEMICS    & Numeric & Share of socially insured employees at the workplace with an academic professional qualification among socially insured employees in percent \\
ACC          & Numeric & Total road traffic accidents per 100,000 inhabitants \\
AGE          & Numeric & Average age of the population in years \\
CAR          & Numeric & Passenger cars per 1,000 inhabitants \\
COM          & Numeric & Commuter balance per 100 socially insured employees at the workplace \\
DEBT         & Numeric & Percentage of private debtors among residents aged 18 and over \\
DIVORCES     & Numeric & Divorces per 1,000 inhabitants aged 18 and over \\
EMPLOYMENT   & Numeric & Percentage of social insurance contributing employees per 100 working-age residents at place of residence \\
FOREIGN      & Numeric & Share of foreigners among the residents in percent \\
GDP          & Numeric & Gross domestic product per inhabitant \\
HHINC        & Numeric & Average household income per inhabitant in euro \\
INS          & Numeric & Consumer insolvency proceedings per 1,000 inhabitants aged 18 and over \\
LABOR        & Numeric & Number of hours worked by employees \\
LAND         & Numeric & Cadastral area in square kilometers \\
LIVE         & Numeric & Living space per inhabitant in square meters \\
MEDINC       & Numeric & Median income of full-time employees subject to social security contributions in euro \\
PART         & Numeric & Labor force participation rate among the working-age population in percent \\
POP          & Numeric & Inhabitants per square kilometer \\
RENT         & Numeric & Rent prices per square meter in euro \\
SELF         & Numeric & Percentage of self-employed among all employed individuals \\
TAX          & Numeric & Tax revenue per inhabitant in euro \\
UNEMPLOYMENT & Numeric & Percentage of unemployed among the civilian labor force \\
WELFARE      & Numeric & Share of employable and non-employable welfare recipients among residents under 65 years old \\
\hline \\[-1.8ex]
\end{tabular}
\end{table}

Thus, to investigate the drivers of life expectancy in German districts from a data-driven variable selection perspective via LTB and DES, a variety of random effects models, namely ANS, KKP, and GSPECM, are considered. For additional comparison, results from classical GMM estimation are also reported. All independent variables presented in Table \ref{tab:var_ger}, their corresponding spatial lags, as well as an intercept, are included in the estimation, leading to a low-dimensional linear setting with $47$ coefficients to estimate in a non-parsimonious model. Furthermore, continuous variables are centered and scaled to ensure comparability across variables and convergence of the LTB algorithm. The setup for LTB and DES closely follows the Monte Carlo experiments in Section \ref{sec:sim}, with the learning rate set to $s = 0.1$ and the threshold to $\tau = 0.01$. The spatial weight matrix is constructed using a 10-nearest-neighbor structure, such that each district is assigned its ten geographically closest neighbors and is subsequently row-normalized. The optimal stopping criterion $m_{\text{opt}}$ is determined by minimizing the empirical risk via 10-fold spatial cross-validation as discussed by \cite{brenning2012}.\footnote{The shapefile of the German districts is provided freely by the Federal Agency for Cartography and Geodesy and can be found at \url{https://gdz.bkg.bund.de/index.php/default/digitale-geodaten/verwaltungsgebiete.html?___store=default}.}

The estimated coefficients for all estimation strategies and different model specifications can be seen in Table \ref{tab:german}. For brevity, the coefficients of the spatial lags of exogenous variables are not reported individually. Instead, the number of spatial lags of variables included in each model (No. $\bm{WX}$) is presented. Regarding the variable selection performance, the results for all considered strategies are comparable. Particularly, LTB reduces the number of variables from $47$ to $36 - 38$ depending on the particular random effects model. The sparsest results are obtained for the ANS random effects model, while the results for the GSPECM are worst, although the range is negligible. Further improvement in variable selection can be seen in DES. Particularly, the number of included variables is further reduced to only $7-8$ across all considered random effects models, where spatial lags of independent variables are not selected.

Estimates of the spatial autocorrelation parameters are negative for $\hat\rho_1$ and positive for $\hat\rho_2$ in the GSPECM, indicating diverging results regarding the persistent spatial dependence between neighboring districts in Germany. In contrast, estimates of the variances of the disturbances $\hat\sigma_{\bm{\mu}}$ and $\hat\sigma_{\bm{\epsilon}}$ remain comparable across all considered random effects models. 

Moreover, the algebraic sign and overall direction of the coefficients remain identical across all estimation strategies. The primary difference lies in the magnitude of the coefficients, which varies due to the shrinkage effects introduced by early stopping in LTB and DES. Taking the results for DES for the GSPECM as an example, the most important variables explaining the life expectancy in German districts in the years $2014-2019$ are the proportion of employees with an academic degree, the average age of the population, the debt quota, the household income, the rent prices, the unemployment rate and the share of welfare recipients. Since the independent variables are transformed by simple scaling and centering, the coefficients have an intuitive and simple interpretation. For instance, an increase in rent prices by one euro ceteris paribus increases the average life expectancy by $0.100$ years on average. Consequently, holding all other variables constant, an one percentage point increase in the share of private debtors is associated with an average $0.368$ year decrease in average life expectancy.

To summarize the findings, the spatial autocorrelation parameters reveal diverging spatial dependence between districts, that is, the disturbances in one district are influenced by the disturbances in neighboring districts in the opposite direction. Furthermore, the share of welfare recipients and average age of the population negatively influence average life expectancy meaning that underlying poor economic and health status decrease the average life expectancy. Furthermore, districts with higher unemployment and debt quotas have on average lower average life expectancies, indicating that financial hardship, fiscal stress or economic precarity is negatively associated with longevity. In contrast, higher rent prices, higher household income and higher share of employees with an academic degree lead to a higher average life expectancy, which shows that inhabitants in wealthier districts and with better socioeconomic status live longer on average \citep{BBSR2024, BKG2025, balzer2025}.

\begin{table}[H]
\caption{Coefficient estimates across different estimation strategies for the German life expectancy data set.} \label{tab:german}
\centering
\setlength{\tabcolsep}{2pt}
\begin{tabular}{lccccccccc}
\\[-1.8ex]\hline
\hline \\[-1.8ex]
 & \multicolumn{3}{c}{GMM} & \multicolumn{3}{c}{LTB} & \multicolumn{3}{c}{DES} \\
\cmidrule{2-4}\cmidrule{5-7}\cmidrule{8-10}
& ANS & KKP & GSPECM 
& ANS & KKP & GSPECM
& ANS & KKP & GSPECM \\
\hline \\[-1.8ex]
$\hat\rho_1$ & 0.000 & 0.212 & -0.284 & 0.000 & 0.212 & -0.284 & 0.000 & 0.212 & -0.284 \\
$\hat\rho_2$ & 0.212 & 0.212 & 0.212 & 0.212 & 0.212 & 0.212 & 0.212 & 0.212 & 0.212 \\
$\hat\sigma_{\bm{\mu}}$ & 0.148 & 0.146 & 0.144 & 0.148 & 0.146 & 0.144 & 0.148 & 0.146 & 0.144 \\
$\hat\sigma_{\bm{\epsilon}}$ & 0.050 & 0.050 & 0.050 & 0.050 & 0.050 & 0.050 & 0.050 & 0.050 & 0.050 \\
(Intercept) & 0.001 & 0.002 & 0.002 & -- & -- & -- & -0.002 & -- & 0.002 \\
ACADEMICS   & 0.284 & 0.284 & 0.283 & 0.250 & 0.250 & 0.250 & 0.179 & 0.181 & 0.184 \\
ACC         & 0.029 & 0.028 & 0.028 & 0.021 & 0.021 & 0.021 & -- & -- & -- \\
AGE         & -0.096 & -0.094 & -0.092 & -0.096 & -0.096 & -0.097 & -0.110 & -0.111 & -0.114 \\
CAR         & 0.057 & 0.056 & 0.057 & 0.037 & 0.036 & 0.036 & -- & -- & -- \\
COM         & -0.048 & -0.048 & -0.048 & -0.049 & -0.042 & -0.042 & -- & -- & -- \\
DEBT        & -0.334 & -0.334 & -0.335 & -0.356 & -0.351 & -0.347 & -0.380 & -0.375 & -0.368 \\
DIVORCES    & 0.024 & 0.024 & 0.025 & 0.023 & 0.023 & 0.023 & -- & -- & -- \\
EMPLOYMENT  & -0.055 & -0.060 & -0.065 & -0.020 & -0.025 & -0.032 & -- & -- & -- \\
FOREIGN     & 0.082 & 0.081 & 0.080 & 0.061 & 0.059 & 0.056 & -- & -- & -- \\
GDP         & -0.101 & -0.102 & -0.101 & -0.092 & -0.093 & -0.091 & -- & -- & -- \\
HHINC       & 0.211 & 0.211 & 0.210 & 0.207 & 0.206 & 0.205 & 0.247 & 0.246 & 0.246 \\
INS         & -0.010 & -0.010 & -0.009 & -0.014 & -0.014 & -0.014 & -- & -- & -- \\
LABOR       & -0.073 & -0.075 & -0.077 & -0.071 & -0.073 & -0.076 & -- & -- & -- \\
LAND        & -0.008 & -0.008 & -0.008 & -0.007 & -0.005 & -0.004 & -- & -- & -- \\
LIVE        & -0.108 & -0.111 & -0.114 & -0.075 & -0.078 & -0.083 & -- & -- & -- \\
MEDINC      & 0.070 & 0.069 & 0.067 & 0.062 & 0.065 & 0.065 & 0.009 & 0.006 & -- \\
PART        & 0.084 & 0.085 & 0.086 & 0.057 & 0.060 & 0.063 & -- & -- & -- \\
POP         & -0.081 & -0.083 & -0.083 & -0.051 & -0.054 & -0.059 & -- & -- & -- \\
RENT        & 0.081 & 0.080 & 0.079 & 0.099 & 0.099 & 0.097 & 0.098 & 0.099 & 0.100 \\
SELF        & 0.056 & 0.058 & 0.060 & 0.043 & 0.048 & 0.051 & -- & -- & -- \\
TAX         & -0.037 & -0.037 & -0.036 & -0.028 & -0.027 & -0.027 & -- & -- & -- \\
UNEMPLOYMENT& -0.076 & -0.079 & -0.084 & -0.088 & -0.087 & -0.087 & -0.048 & -0.051 & -0.055 \\
WELFARE     & -0.093 & -0.090 & -0.087 & -0.060 & -0.063 & -0.066 & -0.112 & -0.109 & -0.107 \\
\hline \\[-1.8ex]
$\text{No.} \bm{WX}$ & 23 & 23 & 23 & 13 & 14 & 15 & 0 & 0 & 0\\
\hline \\[-1.8ex]
\end{tabular}
\begin{minipage}{\textwidth}
\footnotesize
\textbf{Notes:} Reported are the results for the generalized method of moments (GMM), $L_2$-boosting (LTB) and LTB with post-hoc deselection (DES) for \cite{anselin1988} random effects model (ANS), \cite{kapoor2007} random effects model (KKP) and the generalized spatial panel model with error components (GSPECM). For brevity, the coefficients of the spatial lags of exogenous variables are not reported individually. Instead, the number of spatial lag variables included in each strategy (No. $\bm{WX}$) is presented.
\end{minipage}
\end{table}

\section{Conclusion} \label{sec:con}
The key findings and main contributions of this article are: (a) Model-based gradient boosting is extended for the GSPECM, which nests important spatial cases such as the ANS and KKP random effects model. (b) Instead of relying on a spatial error family in practice, a Cochrane–Orcutt type spatial transformation on the data is used, such that model-based gradient boosting reduces to LTB with the $L_2$ loss function. The major advantage of such an approach is the preservation of theoretical properties such as convergence and consistency, which cannot be ensured via the spatial error family. (c) The Cochrane–Orcutt type spatial transformation relies on the knowledge about the spatial autocorrelation parameters $\rho_1$ and $\rho_2$, which are typically unknown in practice, making LTB infeasible. Thus, feasible LTB relies on replacing unknown quantities by estimates based on estimators proposed by \cite{kapoor2007} and \cite{baltagi2016}. (d) Monte Carlo experiments confirm proper functionality of LTB in low- and high-dimensional settings with respect to variable selection and estimation accuracy across varying combinations of spatial autocorrelation parameters and outperform traditional ML and GMM estimators. (e) The application of LTB is illustrated by revisiting classical spatial econometric panel data sets, namely non-life Insurance in Italian provinces, rice production in Indonesian farms and life expectancy in German districts.    

Additionally, limitations, improvements and extensions beyond the scope of this article have to be acknowledged. Although Monte Carlo experiments have been conducted, the scope of the experiments is by no means exhaustive. More complex scenarios such as varying spatial weight matrices or varying data generating process can be considered, which might give a more differentiated view of LTB for the GSPECM. The current implementation of LTB for the GSPECM does not include a spatial lag of the dependent variable. Thus, extending LTB for mixed autoregressive spatial panel models with error components is an interesting avenue for further research. To mitigate consequences arising from low TNR values such as the inclusion of too many non-informative variables, impacting the sparsity of the final model, post-hoc deselection is utilized. However, alternative approaches such as stability selection might be applied to control over the amount of false positive variables \citep{meinshausen2010, shah2013, hofner2015}. Since the Cochrane–Orcutt type spatial transformed GSEPCM is a univariate location model, sparser models can be obtained by applying probing, which stops LTB as soon as the first randomly permuted version of a variable is added \citep{thomas2017}. Additionally, p-values for individual base-learners can be computed by utilizing permutation techniques \citep{hepp2019}. 

Therefore, the next goal is to extend LTB for mixed autoregressive spatial panel models with error components, which is especially challenging in high-dimensional settings due to the increasing model complexity. However, such an extension allows for a more nuanced examination of the empirical illustrations. Furthermore, the development of an efficient software implementation such as a dedicated package in the programming language R is an important avenue for further research. Nevertheless, practitioners and applied statisticians working with spatial panel data are encouraged to utilize LTB for the GSPECM and various nested models such as the ANS and KKP random effects model as a valuable alternative for estimation, regularization, model and variable selection in future research.

\backmatter

\bmhead{Supplementary information}
All R-code for the implemented $L_2$-boosting algorithm along with the Monte Carlo experiments is publicly available in the following GitHub repository \url{https://github.com/micbalz/SpatPanelRegBoost}. 

\section*{Declarations}

\bmhead{Data availability}
The data as well as the accompanying codebook on the life expectancy in German districts is publicly available via \url{https://www.inkar.de/} (DL-DE BY 2.0). The data and the codebook for the non-life insurance in Italian provinces and rice production in Indonesian farms is publicly available in the \texttt{splm} package in R \citep{millo2012, bivand2021}. The shapefile for the North Carolina counties is available in the \texttt{spData} package in R \citep{pebesma2023, bivand2025}.

\bmhead{Funding}
Financial support by the German Research Foundation (DFG) [RTG 2865/1 – 492988838] is gratefully acknowledged.

\bmhead{Conflict of interests}
The authors declare no conflict of interests.

\bibliography{sn-bibliography}

@book{anselin1988,
  title={Spatial econometrics: methods and models},
  author={Anselin, Luc},
  year={1988},
  publisher={Springer Science \& Business Media}
}

@article{kapoor2007,
  title={Panel data models with spatially correlated error components},
  author={Kapoor, Mudit and Kelejian, Harry H and Prucha, Ingmar R},
  journal={Journal of Econometrics},
  volume={140},
  number={1},
  pages={97--130},
  year={2007},
  publisher={Elsevier},
  doi = {10.1016/j.jeconom.2006.09.004}
}

@article{mutl2011,
  title={The {Hausman} test in a {Cliff} and {Ord} panel model},
  author={Mutl, Jan and Pfaffermayr, Michael},
  journal={The Econometrics Journal},
  volume={14},
  number={1},
  pages={48--76},
  year={2011},
  publisher={Oxford University Press Oxford, UK},
  doi = {10.1111/j.1368-423X.2010.00325.x}
}

@article{baltagi2013,
  title={A generalized spatial panel data model with random effects},
  author={Baltagi, Badi H and Egger, Peter and Pfaffermayr, Michael},
  journal={Econometric Reviews},
  volume={32},
  number={5-6},
  pages={650--685},
  year={2013},
  publisher={Taylor \& Francis},
  doi = {10.1080/07474938.2012.742342}
}

@article{baltagi2016,
  title={Random effects, fixed effects and {Hausman's} test for the generalized mixed regressive spatial autoregressive panel data model},
  author={Baltagi, Badi H and Liu, Long},
  journal={Econometric Reviews},
  volume={35},
  number={4},
  pages={638--658},
  year={2016},
  publisher={Taylor \& Francis},
  doi = {10.1080/07474938.2014.998148}
}

@article{xia2023,
  title={Variable Selection of High-Dimensional Spatial Autoregressive Panel Models with Fixed Effects},
  author={Xia, Miaojie and Zhang, Yuqi and Tian, Ruiqin},
  journal={Journal of Mathematics},
  volume={2023},
  number={1},
  pages={9837117},
  year={2023},
  publisher={Wiley Online Library},
  doi = {10.1155/2023/9837117}
}

@article{liu2024,
  title={Adaptive lasso variable selection method for semiparametric spatial autoregressive panel data model with random effects},
  author={Liu, Yu},
  journal={Communications in Statistics-Theory and Methods},
  volume={53},
  number={6},
  pages={2122--2140},
  year={2024},
  publisher={Taylor \& Francis},
  doi = {10.1080/03610926.2022.2119088}
}

@book{baltagi2008,
  title={Econometric analysis of panel data},
  author={Baltagi, Badi Hani},
  year={2008},
  publisher={Springer Cham}
}

@article{lee2010,
  title={Estimation of spatial autoregressive panel data models with fixed effects},
  author={Lee, Lung-fei and Yu, Jihai},
  journal={Journal of econometrics},
  volume={154},
  number={2},
  pages={165--185},
  year={2010},
  publisher={Elsevier},
  doi = {10.1016/j.jeconom.2009.08.001}
}

@article{balzer2025,
  author    = {Balzer, Michael},
  title     = {Gradient Boosting for Spatial Regression Models with Autoregressive Disturbances},
  journal   = {Networks and Spatial Economics},
  year      = {2025},
  doi       = {10.1007/s11067-025-09717-8}
}

@article{bivand2021,
    title = {A Review of Software for Spatial Econometrics in {R}},
    author = {Roger Bivand and Giovanni Millo and Gianfranco Piras},
    journal = {Mathematics},
    year = {2021},
    volume = {9},
    number = {11},
    doi = {10.3390/math9111276},
}

@Article{millo2012,
    title = {{splm}: Spatial Panel Data Models in {R}},
    author = {Giovanni Millo and Gianfranco Piras},
    journal = {Journal of Statistical Software},
    year = {2012},
    volume = {47},
    number = {1},
    pages = {1--38},
    doi = {10.18637/jss.v047.i01},
}

@article{hepp2016,
  author = {Hepp, Tobias and Schmid, Matthias and Gefeller, Olaf and Bergherr, Elisabeth and Mayr, Andreas},
  year = {2016},
  month = {09},
  pages = {422-430},
  title = {Approaches to Regularized Regression - A Comparison between Gradient Boosting and the Lasso},
  volume = {55},
  number = {5},
  journal = {Methods of Information in Medicine},
  doi = {10.3414/ME16-01-0033}
}

@article{mayr2014,
  title={The evolution of boosting algorithms. From machine learning to statistical modelling},
  author={Mayr, Andreas and Binder, Harald and Gefeller, Olaf and Schmid, Matthias},
  journal={Methods of Information in Medicine},
  volume={53},
  number={6},
  pages={419--427},
  year={2014},
  doi = {10.3414/ME13-01-0122}
}

@article{bühlmann2006,
  author = {Peter B{\"u}hlmann},
  title = {{Boosting for high-dimensional linear models}},
  volume = {34},
  journal = {The Annals of Statistics},
  number = {2},
  publisher = {Institute of Mathematical Statistics},
  pages = {559--583},
  year = {2006},
  doi = {10.1214/009053606000000092}
}

@article{bühlmann2007,
  author = {Peter B{\"u}hlmann and Torsten Hothorn},
  title = {Boosting Algorithms: Regularization, Prediction and Model Fitting},
  volume = {22},
  journal = {Statistical Science},
  number = {4},
  publisher = {Institute of Mathematical Statistics},
  pages = {477--505},
  year = {2007},
  doi = {10.1214/07-STS242}
}

@article{friedman2001,
  title={Greedy Function Approximation: A Gradient Boosting Machine},
  author={Friedman, Jerome H},
  journal={Annals of Statistics},
  volume={29},
  number={5},
  pages={1189--1232},
  year={2001},
  publisher={Institute of Mathematical Statistics},
  doi = {10.1214/aos/1013203451}
}

@article{hofner2014,
  title = {Model-based Boosting in {R}: A Hands-on Tutorial Using the {R} Package mboost},
  author = {Benjamin Hofner and Andreas Mayr and Nikolay Robinzonov and Matthias Schmid},
  journal = {Computational Statistics},
  year = {2014},
  volume = {29},
  pages = {3--35},
  doi = {10.1007/s00180-012-0382-5}
}

@article{hothorn2010,
  title = {Model-based Boosting 2.0},
  author = {Torsten Hothorn and Peter Buehlmann and Thomas Kneib and Matthias Schmid and Benjamin Hofner},
  journal = {Journal of Machine Learning Research},
  year = {2010},
  volume = {11},
  pages = {2109--2113},
  doi = {10.5555/1756006.1859922}
}

@Article{hofner2015,
  title = {Controlling false discoveries in high-dimensional situations: Boosting with stability selection},
  author = {Benjamin Hofner and Luigi Boccuto and Markus Goeker},
  journal = {{BMC} Bioinformatics},
  year = {2015},
  volume = {16},
  number = {144},
  doi = {10.1186/s12859-015-0575-3}
}

@manual{R,
    title = {R: A Language and Environment for Statistical Computing},
    author = {{R Core Team}},
    organization = {R Foundation for Statistical Computing},
    address = {Vienna, Austria},
    year = {2026},
    url = {https://www.R-project.org/},
}

@article{zhang2005,
  author = {Tong Zhang and Bin Yu},
  title = {Boosting with early stopping: Convergence and consistency},
  volume = {33},
  journal = {The Annals of Statistics},
  number = {4},
  publisher = {Institute of Mathematical Statistics},
  pages = {1538--1579},
  year = {2005},
  doi = {10.1214/009053605000000255}
}

@article{hao2024,
  title={Model averaging estimation of panel data models with many instruments and boosting},
  author={Hao, Hao and Huang, Bai and Lee, Tae-hwy},
  journal={Journal of applied statistics},
  volume={51},
  number={1},
  pages={53--69},
  year={2024},
  publisher={Taylor \& Francis},
  doi = {10.1080/02664763.2022.2114432}
}

@article{knieper2025,
  title={Gradient boosting for generalised additive mixed models},
  author={Knieper, Lars and Hothorn, Torsten and Bergherr, Elisabeth and Griesbach, Colin},
  journal={Statistics and Computing},
  volume={35},
  number={4},
  pages={1--20},
  year={2025},
  publisher={Springer},
  doi = {10.1007/s11222-025-10612-y}
}

@article{schratz2019,
  title={Hyperparameter tuning and performance assessment of statistical and machine-learning algorithms using spatial data},
  author={Schratz, Patrick and Muenchow, Jannes and Iturritxa, Eugenia and Richter, Jakob and Brenning, Alexander},
  journal={Ecological Modelling},
  volume={406},
  pages={109--120},
  year={2019},
  publisher={Elsevier},
  doi = {10.1016/j.ecolmodel.2019.06.002}
}

@inproceedings{brenning2012,
  title={Spatial cross-validation and bootstrap for the assessment of prediction rules in remote sensing: The {R} package sperrorest},
  author={Brenning, Alexander},
  booktitle={2012 IEEE international geoscience and remote sensing symposium},
  pages={5372--5375},
  year={2012},
  organization={IEEE},
  doi = {10.1109/IGARSS.2012.6352393}
}

@article{valavi2019,
  author = {Valavi, Roozbeh and Elith, Jane and Lahoz-Monfort, José J. and Guillera-Arroita, Gurutzeta},
  title = {{BlockCV}: An {R} package for generating spatially or environmentally separated folds for k-fold cross-validation of species distribution models},
  journal = {Methods in Ecology and Evolution},
  volume = {10},
  number = {2},
  pages = {225-232},
  year = {2019},
  doi = {10.1111/2041-210X.13107}
}

@article{schratz2024,
    title = {{mlr3spatiotempcv}: Spatiotemporal Resampling Methods for
      Machine Learning in {R}},
    author = {Patrick Schratz and Marc Becker and Michel Lang and
      Alexander Brenning},
    journal = {Journal of Statistical Software},
    year = {2024},
    volume = {111},
    number = {7},
    pages = {1--36},
    doi = {10.18637/jss.v111.i07}
}

@article{stromer2022,
  title={Deselection of base-learners for statistical boosting—with an application to distributional regression},
  author={Str{\"o}mer, Annika and Staerk, Christian and Klein, Nadja and Weinhold, Leonie and Titze, Stephanie and Mayr, Andreas},
  journal={Statistical Methods in Medical Research},
  volume={31},
  number={2},
  pages={207--224},
  year={2022},
  publisher={SAGE Publications Sage UK: London, England},
  doi = {10.1177/0962280221105108}
}

@article{schmid2008,
  author = {Schmid, Matthias and Hothorn, Torsten},
  title = {Boosting additive models using component-wise {P-Splines}},
  journal = {Computational Statistics \& Data Analysis},
  volume = {53},
  number = {2},
  pages = {298--311},
  year = {2008},
  doi = {10.1016/j.csda.2008.09.009}
}

@article{mayr2012,
  title={The importance of knowing when to stop},
  author={Mayr, Andreas and Hofner, Benjamin and Schmid, Matthias},
  journal={Methods of Information in Medicine},
  volume={51},
  number={02},
  pages={178--186},
  year={2012},
  doi = {10.3414/ME11-02-0030}
}

@article{stehmann1997,
  title = {Selecting and interpreting measures of thematic classification accuracy},
  journal = {Remote Sensing of Environment},
  volume = {62},
  number = {1},
  pages = {77--89},
  year = {1997},
  issn = {0034-4257},
  author = {Stephen V. Stehman},
  doi = {10.1016/S0034-4257(97)00083-7}
}

@article{morris2019,
  title={Using simulation studies to evaluate statistical methods},
  author={Morris, Tim P and White, Ian R and Crowther, Michael J},
  journal={Statistics in Medicine},
  volume={38},
  number={11},
  pages={2074--2102},
  year={2019},
  publisher={Wiley Online Library},
  doi = {10.1002/sim.8086}
}

@article{millo2011,
  title={Non-life insurance consumption in Italy: a sub-regional panel data analysis},
  author={Millo, Giovanni and Carmeci, Gaetano},
  journal={Journal of Geographical Systems},
  volume={13},
  number={3},
  pages={273--298},
  year={2011},
  publisher={Springer},
  doi = {10.1007/s10109-010-0125-5}
}

@article{millo2014,
  title={Maximum likelihood estimation of spatially and serially correlated panels with random effects},
  author={Millo, Giovanni},
  journal={Computational Statistics \& Data Analysis},
  volume={71},
  pages={914--933},
  year={2014},
  publisher={Elsevier},
  doi = {10.1016/j.csda.2013.07.024}
}

@article{millo2022,
  title={The generalized spatial random effects model in R},
  author={Millo, Giovanni},
  journal={Journal of Spatial Econometrics},
  volume={3},
  number={1},
  pages={7},
  year={2022},
  publisher={Springer},
  doi = {10.1007/s43071-022-00024-9}
}

@article{druska2004,
  title={Generalized moments estimation for spatial panel data: Indonesian rice farming},
  author={Druska, Viliam and Horrace, William C},
  journal={American Journal of Agricultural Economics},
  volume={86},
  number={1},
  pages={185--198},
  year={2004},
  publisher={Wiley Online Library}, 
  doi = {10.1111/j.0092-5853.2004.00571.x}
}

@article{horrace1996,
  title={Confidence statements for efficiency estimates from stochastic frontier models},
  author={Horrace, William C and Schmidt, Peter},
  journal={Journal of Productivity Analysis},
  volume={7},
  number={2},
  pages={257--282},
  year={1996},
  publisher={Springer},
  doi = {10.1007/BF00157044}
}

@article{horrace2000,
  title={Multiple comparisons with the best, with economic applications},
  author={Horrace, William C and Schmidt, Peter},
  journal={Journal of Applied Econometrics},
  volume={15},
  number={1},
  pages={1--26},
  year={2000},
  publisher={Wiley Online Library},
  doi = {10.1002/(SICI)1099-1255(200001/02)15:1<1::AID-JAE551>3.0.CO;2-Y}
}

@misc{BKG2025,
  author={{Bundesamt für Kartographie und Geodäsie (BKG)}},
  title={Geodaten des BKG},
  year={2024},
  url={https://www.bkg.bund.de}
}

@misc{BBSR2024,
  author={{Bundesinstitut für Bau-, Stadt- und Raumforschung (BBSR)}},
  title={{Laufende Raumbeobachtung des BBSR - INKAR, Ausgabe 03/2024}},
  year={2024},
  url={https://www.inkar.de/}
}

@article{meinshausen2010,
  author = {Meinshausen, Nicolai and Bühlmann, Peter},
  title = {Stability Selection},
  journal = {Journal of the Royal Statistical Society Series B: Statistical Methodology},
  volume = {72},
  number = {4},
  pages = {417-473},
  year = {2010},
  month = {08},
  doi = {10.1111/j.1467-9868.2010.00740.x},
}

@article{shah2013,
  author = {Shah, Rajen D. and Samworth, Richard J.},
  title = {Variable Selection with Error Control: Another Look at Stability Selection},
  journal = {Journal of the Royal Statistical Society Series B: Statistical Methodology},
  volume = {75},
  number = {1},
  pages = {55-80},
  year = {2012},
  month = {06},
  doi = {10.1111/j.1467-9868.2011.01034.x},
}

@article{thomas2017,
  title={Probing for sparse and fast variable selection with model-based boosting},
  author={Thomas, Janek and Hepp, Thomas and Mayr, Andreas and Bischl, Bernd},
  journal={Computational and Mathematical Methods in Medicine},
  volume={2017},
  year={2017},
  month = {07},
  publisher={Hindawi},
  doi = {10.1155/2017/1421409}
}

@article{hepp2019,
  title = {Significance Tests for Boosted Location and Scale Models with Linear Base-Learners},
  author = {Tobias Hepp and Matthias Schmid and Andreas Mayr},
  volume = {15},
  number = {1},
  journal = {The International Journal of Biostatistics},
  year = {2019},
  doi = {10.1515/ijb-2018-0110}
}

@Book{pebesma2023,
  author = {Edzer Pebesma and Roger S. Bivand},
  title = {Spatial Data Science With Applications in {R}},
  year = {2023},
  publisher = {Chapman \& Hall},
  url = {https://r-spatial.org/book/}
}

@Manual{bivand2025,
  title = {spData: Datasets for Spatial Analysis},
  author = {Roger Bivand and Jakub Nowosad and Robin Lovelace},
  year = {2025},
  note = {R package version 2.3.4},
  url = {https://jakubnowosad.com/spData/},
}

\end{document}